\documentclass[aps,prx,twocolumn,superscriptaddress]{revtex4-2}
\usepackage[a4paper,margin=1.25cm]{geometry}
\usepackage{times}
\usepackage{graphicx}
\usepackage[font={small}]{caption}
\usepackage{subcaption}
\usepackage{amsmath}
\usepackage{amssymb}
\usepackage{tabularx}
\usepackage{placeins}
\usepackage{units}
\usepackage{multirow}
\usepackage{braket}
\usepackage{color}
\usepackage{hyperref}
\usepackage[normalem]{ulem}
\usepackage{breakcites}
\usepackage{adjustbox}
\usepackage{mathtools}
\usepackage{float}
\usepackage{adjustbox}
\usepackage{dcolumn}
\usepackage{bm}
\usepackage{comment}
\usepackage{amssymb}

\usepackage[dvipsnames]{xcolor}

\begin{document}

\title{A new generation of effective core potentials: Selected lanthanides and heavy elements II}

\author{Omar Madany}
\email{osmadany@ncsu.edu}
\author{Benjamin Kincaid} 
\author{Aqsa Shaikh}
\author{Elizabeth Morningstar}
\author{Lubos Mitas}
\affiliation{
Department of Physics, North Carolina State University, Raleigh, North Carolina 27695-8202, USA \\
}

\begin{abstract}
We present a new set of correlation-consistent effective core potentials (ccECPs) for selected heavy $s$, $p$, $d$, and $f$-block elements significant in materials science and chemistry (Rb, Sr, Cs, Ba, In, Sb, Pb, Ru, Cd, La, Ce, and Eu). The ccECPs are designed using minimal Gaussian parameterization to achieve smooth and bounded potentials. They are expressed as a combination of averaged relativistic effective potentials (AREP) and effective spin-orbit (SO) terms, developed within a relativistic coupled-cluster framework.
The optimization is driven by correlated all-electron (AE) atomic spectra, norm-conservation, and spin-orbit splittings, with considerations for plane wave cut-offs to ensure accuracy and viability across various electronic configurations. The transferability of these ccECPs is validated through testing on molecular oxides and hydrides, emphasizing discrepancies in molecular binding energies across a spectrum of bond lengths and electronic environments. The ccECPs demonstrate excellent agreement with AE reference calculations, attaining chemical accuracy in bond dissociation energies and equilibrium bond lengths, even in systems characterized by substantial relativistic and correlation effects. These ccECPs provide an accurate and transferable framework for valence-only calculations.
\end{abstract}

\maketitle

\section{Introduction}
Effective core potentials (ECPs) and closely related pseudopotentials are essential tools in electronic structure calculations, enabling us to focus on valence electrons by eliminating core electrons and the corresponding core energy scales. This partitioning assumes that cores have only marginal influence on the properties of valence electronic states that determine behavior of chemical and condensed matter systems such as bonding, reactivity, magnetic effects, and many others. In an ECP construction, the core states are substituted by potentials that mimic the original effects of the core on the valence electrons, significantly simplifying the subsequent valence-only calculations.
This is particularly important for heavy elements from the fifth and sixth rows of the periodic table, where core states modifications due to relativistic effects complicate the electron-electron correlation effects making thus all-electron (AE) calculations more challenging. In such scenarios, ECPs become almost indispensable as they improve efficiency while effectively capture major parts of core-valence and core-core correlations that could otherwise be neglected, for example, in frozen core approaches.

Notably, over recent years, we have developed correlation-consistent ECPs (ccECPs), enhancing their overall fidelity to AE Hamiltonians through correlated many-body methodologies applied consistently to both AE and ECP construction~\cite{Bennett2017,Bennett2018,Annaberdiyev2018,Wang2019,Wang2022,Kincaid2022,Haihan2024}. Primary objectives include reproducing valence many-body atomic energy differences (excitations and ionizations), ensuring finite wavefunction cusp conditions at the origin, and improving transferability. The latter has been achieved by matching binding curves of hydride and oxide molecules to their fully correlated AE counterparts, thereby probing covalent and ionic bonding. We have concurrently explored the performance limits of simple semi-local Gaussian-based forms, which enable convenient implementation across computational frameworks. The reliability of ccECPs has been rigorously established through extensive validation both in our efforts and by independent research groups, who have consistently leveraged its pragmatism in diverse many-body calculations—from solids and large molecular complexes to neural-network-based quantum simulations—demonstrating robust transferability across chemical environments\cite{Annaberdiyev2020, Annaberdiyev2021,Annaberdiyev2022,Melton2020,Wang2021,Li2022,Wines2020,Oqmhula2020,Otis2020,Wang2019a,Zhou2019,Tyagi2023,Shin2021,Hill2022,Goldzak2022,Denawi2023,Wang2024}.

In contrast, legacy ECPs (referring to previous-generation constructions) were typically built around self-consistent approaches such as Hartree-Fock/Dirac-Fock (HF/DF) and a variety of Density Functional Approximations, focusing on reproducing one-particle atomic eigenvalues, norm/charge conservation outside the cores, and total energy differences between ground and excited states \cite{BFD-2007,BFD-2008,SDFSTU-1982,SDFSTU-1983,MDFSTU-2000,MDFSTU-2005,MDFSTU-2006,MDFSTU-2007,MDFSTU-K-119,MWBSTU-1990,MWBSTU-1991-Ca-Ba,MWBSTU-1991-Hg-Rn,MWBSTU-1993,MWBSTU-1996-1,MWBSTU-1996-2,MWBSTU-1997,MWBSTU-Ce,MWBSTU-La,SDFSTU-1982,SDFSTU-1983,MHFSTU-La,MWBSTU-1991-Ca-Ba,CREN-1987,CREN-1990,SBKJC-1992,SBKJC-1993,LANL2-1985-1,LANL2-1985-2}.

Our previous careful testing revealed and quantified also systematic errors for cases with large shallow cores and resulting small number of valence electrons for example, for Na[Ne] and Mg[Ne] (here the element in the bracket implies the size of the ionic core). Occasionally, we have found significant discrepancies with regard to AE references and for these cases we tested smaller cores such as Na[He] which showed excellent fidelity with AE results. Further developments have included 
Kleinman-Bylander transformation for use in plane wave codes, softening the ECPs for lower plane wave cut-offs in codes with periodic boundary conditions, and providing the basis sets appropriate for correlated calculations.

In this continuing work, we expand ccECP library to include elements specifically requested by the user community through the pseudopotential library\cite{pseudopotentiallibrary}: $5s$ and $6s$ metals (Rb, Sr, Cs, Ba), $5p$ and $6p$ post-transition metals and metalloids (In, Sb, Pb), as well as technologically important transition metals (Ru, Cd) and $4f$ lanthanides (La, Ce, Eu). This community-directed expansion addresses urgent computational needs across frontier applications including heterogeneous catalysis, advanced materials design, and correlated electron systems in condensed matter physics. Maintaining our methodology's core principles, we employ minimal parameterization with simple Gaussian functional forms to ensure smooth, bounded potentials. Transferability is further verified through rigorous validation across diverse bonding environments, including oxide and hydride binding curves that span comprehensive bond-length regimes.

Our ccECPs provide a significant step forward in achieving accuracy, consistency, and an priori understanding of residual systemtic errors for the valence-only effective Hamiltonian model. By employing scalar relativistic effects and incorporating core-valence correlations from the outset, we establish a framework that offers the desired accuracy for valence calculations and facilitates broader transferability across multiple computational methods and systems. 

The following sections describe construction methodologies and objective functions, present the details and results of optimizations, and discuss the implications of our findings. Subsection \ref{ECP form} outlines the theoretical foundation underlying this work, while subsection \ref{sec:optpro} outlines the optimization procedure employed. In Section \ref{Results} we define the metrics utilized for evaluating the performance of the ECPs and provide a comprehensive overview of their effectiveness across all elements studied herein. Subsections \ref{Selected 5s elements} through \ref{Selected 4f elements} give short detailed insights into each element, accompanied by the parameters of the developed ccECPs, as presented in Tables \ref{tab:selected_5s_params} to \ref{tab:selected_4f_params}. Finally, we summarize our findings in Section \ref{Conclusions}.

\section{Methods}
\label{Methods}
\subsection{ECP form}
\label{ECP form}

We approximate the relativistic AE Hamiltonian with a simpler, valence-only Hamiltonian ($H_{\mathit{val}}$), and we always assume the Born-Oppenheimer approximation\cite{Born-Oppenheimer-1927} in all calculations. In atomic units $H_{\mathit{val}}$ is expressed as follows:
\begin{equation}
    H_{\mathit{val}} = \sum_{i} \left[ T^{\text{kin}}_{i} + V^{\text{SOREP}}_{i} \right] + \sum_{i<j} \frac{1}{r_{ij}}
\end{equation}
where $T^{\text{kin}}_i$ is the kinetic energy of the $i$-th electron, and $V^{\text{SOREP}}_i$ represents the semi-local, two-component spin–orbit relativistic ECP (SOREP). 
The general form of $V^{\text{SOREP}}_i$, as developed in Refs.\cite{Lee-1977,Ermler1981}, is expressed as follows:
\begin{equation}\begin{aligned}
    V^{\text{SOREP}}_i &= V_L(r_i) 
    + \sum_{\ell=0}^{\ell_{\text{max}} = L-1} \sum_{j=|\ell-\frac{1}{2}|}^{\ell+\frac{1}{2}} \sum_{m=-j}^{j} \\
    &\quad \times \left[V^{\text{SOREP}}_{\ell j}(r_i) - V_L(r_i)\right] 
    \lvert \ell jm \rangle \langle \ell jm \rvert,
\end{aligned}\end{equation}
here $r_i$  represents the distance of the $i$-th electron from the core’s origin, and $V_L$ represents the local potential of $V^{\text{SOREP}}_i$ and $L$ is selected to be one higher than the maximum angular quantum number of the core electrons ($\ell_{\text{max}}$).
The difference $V^{\text{SOREP}}_{\ell j} - V_L$ accounts for the non-local corrections to the local potential channels of $V^{\text{SOREP}}_i$.
The projection operators $\lvert \ell jm \rangle \langle \ell jm \rvert$ are two-component angular basis functions, which are eigenfunctions in the Pauli approximation of the Dirac Hamiltonian. 
These operators ensure that $V^{\text{SOREP}}_{\ell j}$ acts only on spinors with the correct angular symmetry allowing for the incorporation of relativistic quantum states defined by both spin-orbit coupling and orbital angular momentum.\\

Initially, the spin-orbit effects in $V^{\text{SOREP}}_i$ are averaged out, yielding the averaged relativistic effective potential (AREP), $V^{\text{AREP}}_i$. 
In this case, the projection operators now act only on subspaces characterized by $\ell$ and $m$ only. 
The $V^{\text{AREP}}_i$ operator, therefore, captures all relativistic effects except explicit spin-orbit coupling:
\begin{equation}\begin{aligned}\label{Varepform}
    V^{\text{AREP}}_i &= V_L(r_i) 
    + \sum_{\ell=0}^{\ell_{\text{max}}=L-1} \left( V_\ell(r_i) - V_L(r_i) \right) \\
    &\quad \times \sum_{m=-\ell}^{\ell} | \ell m \rangle \langle \ell m |,
\end{aligned}\end{equation}
such that $V_\ell - V_L$ refers to the non-local portion of $V^{\text{AREP}}_i$.\\ 

The spin-orbit coupling effects are recovered by taking the difference between $V^{\text{SOREP}}_i$ and $V^{\text{AREP}}_i$, which involves expanding the projection operators in terms of two-component spinor operators.
Hence, $V^{\text{SOREP}}_i$ can be understood as consisting of two terms: the AREP and the spin-orbit (SO) contribution.
\begin{equation}\label{eq:sorep_arep_so}
    V^{\text{SOREP}}_i = V^{\text{AREP}}_i + V^{\text{SO}}_i,
\end{equation}
with $V^{\text{SO}}_i$ is defined here as:
\begin{equation}
    V^{\text{SO}}_i = \sum_{\ell=1}^{\ell_{\text{max}}} \Delta V_{\ell}^{\text{SO}} \, P_{\ell} \, \vec{\ell} \cdot \vec{s} \, P_{\ell}
\end{equation}
In this expression, $\Delta V_{\ell}^{\text{SO}}$ the spin–orbit radial potential, and $P_{l}$ the projection operators described earlier in Eq.(\ref{Varepform}).\\

In the ccECPs we define the local potential $V_L$ in a way that cancels the Coulomb singularity, reducing the sharpness of the potential at small $r_i$ and smoothing its curvature\cite{BFD-2007,BFD-2008}:
\begin{equation}\label{eq:AREP_local}
    V_L (r_i) = - \frac{Z_{\text{eff}}}{r} \left( 1 - e^{\alpha r^2} \right) + \alpha Z_{\text{eff}} r e^{-\beta r^2} + \sum_{i=1}^{2} \gamma_i e^{-\delta_i r^2}    
\end{equation}
in which $Z_{\text{eff}}$ is the effective core charge representing the number of valence electrons, and $\alpha$, $\beta$, $\delta_i$, and $\gamma_i$ are parameters for the local potential optimization. 
The non-local potential is defined as:
\begin{equation}\label{eq:AREP_non-local}
    V_\ell(r_i) - V_L(r_i) = \sum_{k=1} \beta_{\ell k} r^{n_{\ell k} - 2} e^{-\alpha_{\ell k} r^2}
\end{equation}
where $n_{\ell k}$ are fixed integers, typically set to 2 before optimization.
The parameters $\alpha_{\ell k}$ and $\beta_{\ell k}$ are optimized according to Sec.\ref{sec:optpro}. 
The spin–orbit radial potential $\Delta V_{\ell}^{\text{SO}}$ follows a similar form:
\begin{equation}\label{eq:SO}
    \Delta V_{\ell}^{\text{SO}} = \sum_{k} \beta_{\ell k} r^{n_{\ell k} - 2} e^{-\alpha_{\ell k} r^2}
\end{equation}
Analogously, $n_{\ell k}$ are fixed integers, and $\alpha_{\ell k}$ and $\beta_{\ell k}$ are parameters for optimization.

\subsection{Optimization procedure}
\label{sec:optpro}
In order to find the optimal values for the Gaussian parameters denoted with Greek letters in Eqs.(\ref{eq:AREP_local}--\ref{eq:SO}) we follow an optimization process that begins with constructing the AREP portion, as this component is essential for achieving the required accuracy and transferability of the ccECP.
Once this foundation is established, we proceed to optimize the SO terms, ensuring that the full relativistic effects are accurately incorporated into the model.

The initial step of AREP construction is selecting the target element and core configuration (detailed per element in Sec.\ref{sec:results}). 
Initially, we employ a numerical HF/DF code\cite{yoon-DHF,Kotochigova1997}, to ensure that the used basis set and sometimes the effects of lower symmetries such as $D_{2h}$ in Gaussian codes do not affect the comparison of the states of interest in both the AE and the ECP settings. These aspects are checked by verifying that the energy gaps from the numerical and Gaussian codes agree with high precision for every state that is tested in the atomic spectrum.
In particular, we minimize errors related to our very large uncontracted basis sets\cite{basis-2007,basis-2016,basis-2018,BSE} (e.g., aug-cc-pwCVTZ), by eliminating near-linear dependencies and adding diffuse terms as needed.
The tested atomic spectrum encompasses a range of energy levels, electron affinities (EA), various excited states, and ionization potentials (IPs), all of which are critical for accurately capturing the effective charge states typically encountered in molecular and solid-state environments.
Then we perform coupled cluster calculations with singles, doubles, and perturbative triples (CCSD(T)) to obtain the atomic spectra for the AE and ECP cases.
All CCSD(T) calculations used Molpro\cite{molpro-2012,molpro-2020,molpro-3,molpro-cc,molpro-seward,molpro-mcscf1,molpro-mcscf2} version 2024.2, with AE calculations specifically employing the tenth-order Douglas-Kroll-Hess Hamiltonian\cite{DKH-2004} to account for scalar relativistic effects. \\

The initial AREP parameters are often guessed from the Stuttgart ECPs or prior ccECPs, with additional Gaussian functions added to the local channel to remove the Coulomb cusp as shown in Eq.(\ref{eq:AREP_local}).
For parameter optimization, we have used the nonlinear optimization code DONLP2\cite{Spellucci-1998,Spellucci-1998-sqp,Spellucci-2009} which employs a sequential quadratic programming algorithm.
This technique is particularly well-suited to our problem as it effectively handles the non-linearities inherent in the parameterization of ECPs by considering second derivatives of both the objective functions and constraints.
Given the smoothness constraints applied to the ccECPs parameters, the solver's ability to manage such nonlinear optimization problems is especially beneficial.\\

The objective function in Eq.(\ref{eq:AREP-obj_fxn}) minimizes discrepancies between AE atomic spectrum gaps and corresponding the AREP gaps. Moreover, it prioritizes the reproduction of the lowest single-particle eigenvalues in the valence, which are vital for characterizing the tails of the one-particle states and ensuring charge/norm conservation. The objective function is formalized as:
\begin{equation}\label{eq:AREP-obj_fxn}
    \mathcal{O}^{2}_{\text{AREP}} = \sum_{s \in S} \omega_s \left( \Delta E^s_{\text{ECP}} - \Delta E^s_{\text{AE}} \right)^2 + \sum_{i \in I} \omega_i \left( \epsilon^i_{\text{ECP}} - \epsilon^i_{\text{AE}} \right)^2,
\end{equation}
where $\Delta E^s$ denotes the atomic energy gap for state $s$ relative to the ground state, and $\epsilon^i$ correspond to the single-particle eigenvalue, applicable in both the ECP and AE contexts.
The first term in the function quantifies the sum of squared biases among the selected states from the AE atomic spectrum, with corresponding weights $\omega_s$ that allow for flexibility in emphasizing particular states, thereby improving optimization convergence in a number of cases.
The optimized AREP parameters are then used to calculate the atomic spectrum at the CCSD(T) level, and the results are compared with those from legacy ECPs.\\

Additionally, ccECPs are examined in diatomic molecular systems, and the process is repeated until the desired accuracy is achieved, approximately within chemical accuracy (1 kcal/mol $\approx$ 0.043 eV).
In the selected $4f$ elements (Sec.\ref{Selected 4f elements}), to converge to the corresponding state between the AE and the ECP cases, we had to carefully construct the initial orbitals by utilizing MOLPRO's \texttt{MERGE} routine in Refs.\cite{molpro-2012, molpro-2020, molpro-3}. 
We targeted the same bonding configuration prescribed by Dolg and coworkers in their work on the lanthanides\cite{MHFSTU-La}.
Skipping this step, the AE and ECPs converged to slightly different states, due to the degeneracy of $4f$, $5d$ and $6s$ orbitals, leading to inscrutable binding discrepancies that showed no specific trend.
To make use of this procedure, we tested the monohydrides as they were simpler to configure and converge than the trihydrides that were tested in our previous work\cite{Haihan2024}, while still targeting covalent type bonding to gauge performance.
Due to the complexity present in the molecular systems for the selected $4f$ elements, sole reliance on the optimizer was found to bias the ECP towards the atomic spectrum. 
Consequently, we initiated the optimization procedure from a moderately performative ECP and carefully tuned the parameters monitoring the performance in the atomic spectrum and the molecular binding curves.\\

Following the optimization of the AREP ECPs and subsequent transferability assessments, we proceeded to optimize the spin–orbit terms. The objective function for the spin–orbit terms can similarly be articulated as:
\begin{equation}
    \mathcal{O}^{2}_{\text{SO}} = \sum_{s \in S} \omega_s \left( \Delta E^s_{\text{ECP}} - \Delta E^s_{\text{AE}} \right)^2 + \sum_{m \in M} \omega_m \left( \epsilon^m_{\text{ECP}} - \epsilon^m_{\text{AE}} \right)^2,
\end{equation}
with a critical distinction being the focus on multiplet splitting energies. These energies are referenced from the lowest energy within the same electronic state and denoted by $\epsilon^m$.

All calculations pertaining to SO terms optimizations were performed using the DIRAC code\cite{DIRAC22,DIRAC22-2020} at the complete open-shell configuration interaction (COSCI) level of accuracy, employing the exact two-component (X2C) Hamiltonian\cite{DIRAC-X2C, DIRAC-X2C-2}.\\

Upon finalizing the constructions of the AREP and SO contributions, we proceed to generate even-tempered (aug)-cc-p(C)VnZ basis sets, optimizing the lowest exponent and the ratio between successive exponents for each basis channel. This procedure was complemented by the inclusion of augmentation terms. Further elaboration on basis set generation detailed in the supplementary material.

\section{Results}\label{Results}\label{sec:results}
Similarly to our previous studies, we used three primary metrics to evaluate the accuracy of the atomic spectrum. The first is the mean absolute deviation (MAD), which assesses the deviation of ECP-calculated energy gaps for $N$ states relative to the corresponding AE reference gaps, defined as follows:
\begin{equation}
    \text{MAD} = \frac{1}{N} \sum_{s=1}^{N} \left| \Delta E^{\text{ECP}}_s - \Delta E^{\text{AE}}_s \right|
\end{equation}

\begin{figure}[!htbp]
\centering
\includegraphics[width=1.00\columnwidth]{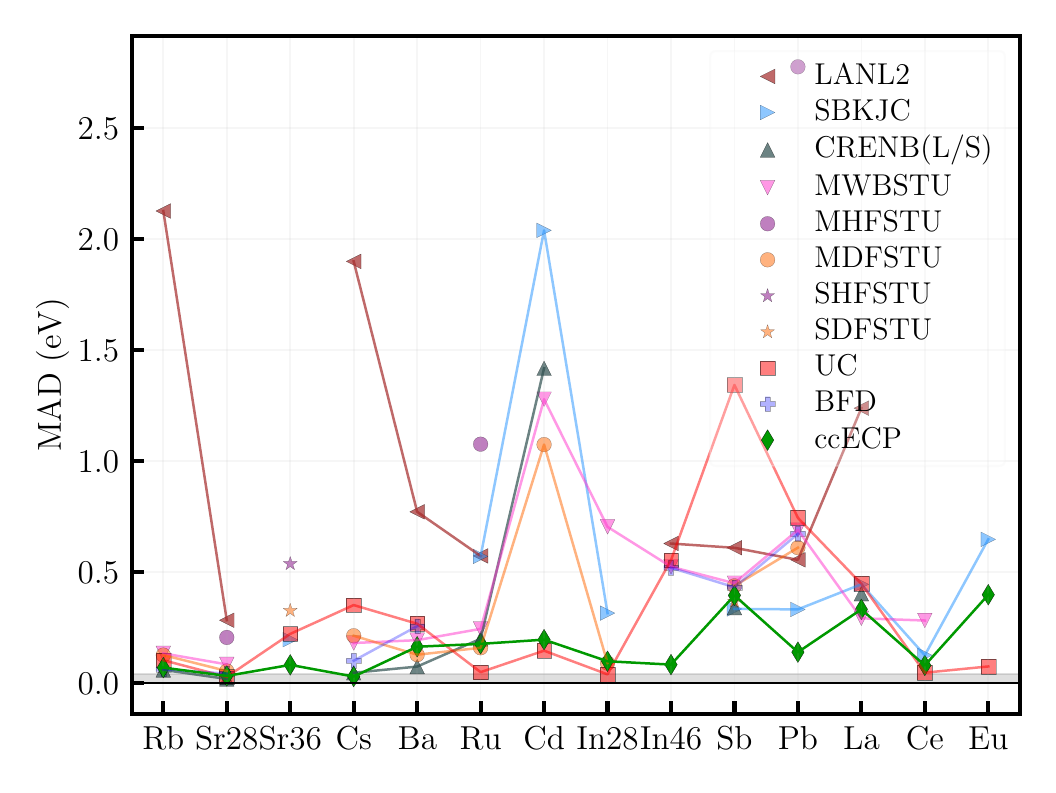}
\caption{
Scalar relativistic AE gap MADs for various core approximations using RCCSD(T) method with the exception of Pb, which was calculated using UCCSD(T). The number next to the element label shows the number of electrons in the core.  
}
\label{fig:MAD_in_elements}
\end{figure}

The second metric focuses on a subset $n$ of low-lying states, including electron affinity (EA) (if a stable anion exists for the given element), first ionization potential (IP), second ionization potential (IP2), and the first excited states; this is referred to as the low-lying mean absolute deviation (LMAD) and is given by:
\begin{equation}
    \text{LMAD} = \frac{1}{n} \sum_{s=1}^{n} \left| \Delta E^{\text{ECP}}_s - \Delta E^{\text{AE}}_s \right|
\end{equation}

\begin{figure}[!htbp]
\centering
\includegraphics[width=1.00\columnwidth]{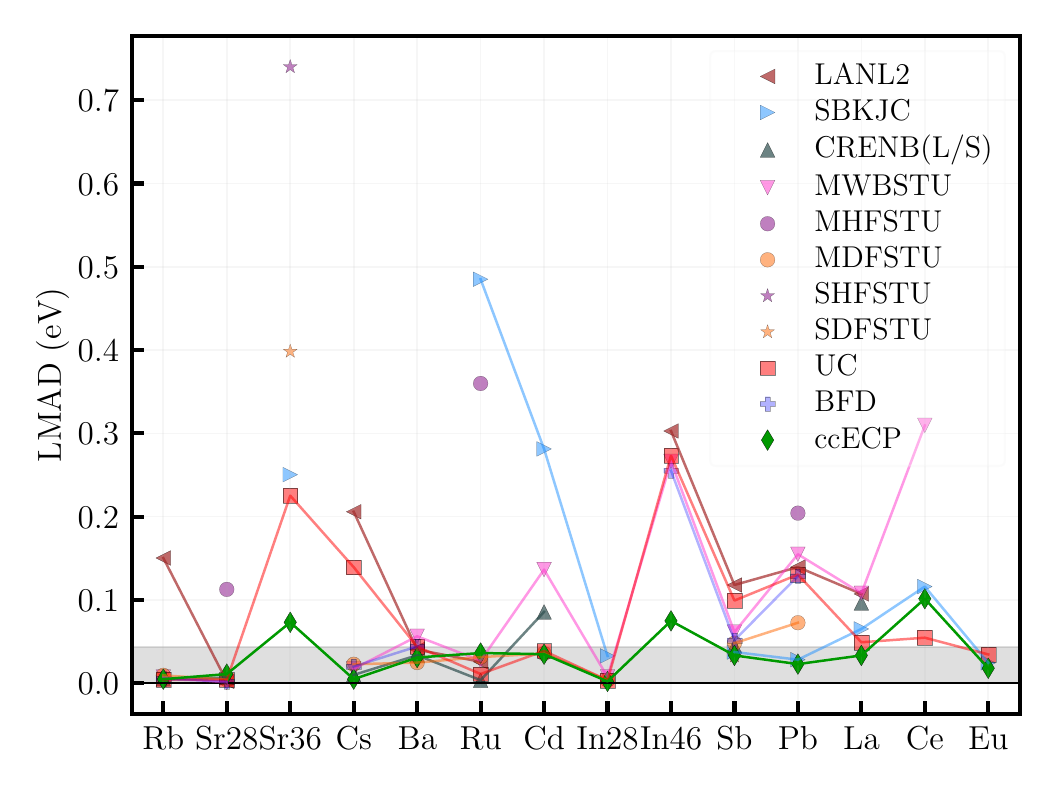}
\caption{
Scalar relativistic AE gap LMADs for various core approximations (Method details same as Fig. \ref{fig:MAD_in_elements}). 
}
\label{fig:LMAD_in_elements}
\end{figure}

Lastly, we introduce the weighted mean absolute deviation (WMAD) to give a percentage-like perspective for larger errors present for gaps with large  magnitudes, as expressed by:
\begin{equation}
    \text{WMAD} = \frac{1}{N} \sum_{s=1}^{N} \frac{100\%}{\sqrt{\left| \Delta E^{\text{AE}}_s \right|}}  \left| \Delta E^{\text{ECP}}_s - \Delta E^{\text{AE}}_s \right|
\end{equation}

\begin{figure}[!htbp]
\centering
\includegraphics[width=1.00\columnwidth]{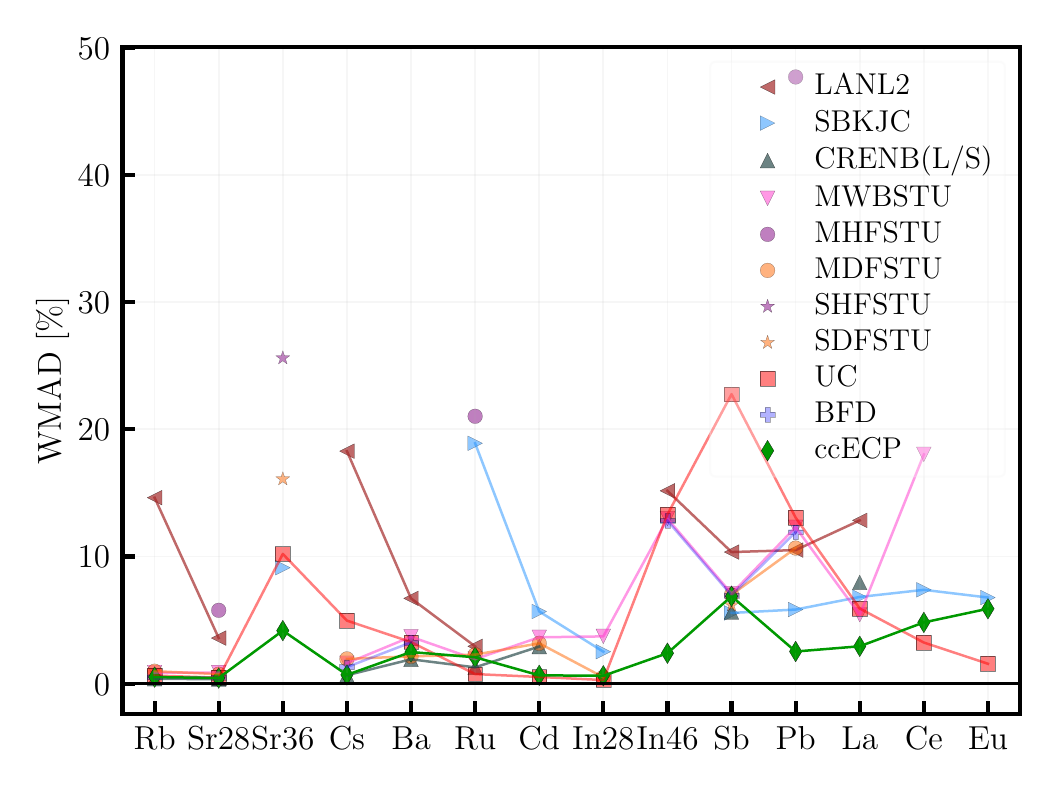}
\caption{
Scalar relativistic AE gap WMADs for various core approximations (Method details same as Fig. \ref{fig:MAD_in_elements}). 
}
\label{fig:WMAD_in_elements}
\end{figure}

To benchmark our ccECP constructions, we compute the same metrics for several legacy ECPs, including BFD\cite{BFD-2007,BFD-2008}, (S/M)DFSTU\cite{SDFSTU-1982,SDFSTU-1983,MDFSTU-2000,MDFSTU-2005,MDFSTU-2006,MDFSTU-2007,MDFSTU-K-119}, MWBSTU\cite{MWBSTU-1990,MWBSTU-1991-Ca-Ba,MWBSTU-1991-Hg-Rn,MWBSTU-1993,MWBSTU-1996-1,MWBSTU-1996-2,MWBSTU-1997,MWBSTU-Ce,MWBSTU-La}, (S/M)HFSTU\cite{SDFSTU-1982,SDFSTU-1983,MHFSTU-La,MWBSTU-1991-Ca-Ba}, CRENB(S/L)\cite{CREN-1987,CREN-1990}, SBKJC\cite{SBKJC-1992,SBKJC-1993}, and LANL2\cite{LANL2-1985-1,LANL2-1985-2}, if they contain ECPs with cores similar to the ccECP for the element being targeted.
Additionally, we include uncorrelated-core (UC) all-electron calculations, where the valence space is fully correlated while the core is frozen, in order to compare with another version of an effective core setting.
The results for MAD, LMAD, and WMAD across all elements are summarized graphically in Figs.\ref{fig:MAD_in_elements}-\ref{fig:WMAD_in_elements}, illustrating the improved accuracy and consistency of the ccECPs relative to the legacy ECPs.
We note that, CRENB(S/L), SBKJC, and STU ECP parameterization differ from ccECP and BFD, which are specifically designed to eliminate any Coulomb cusps and provide potentials which are bounded everywhere.
Additionally, during the optimization process, we restricted the range of all exponents ($\alpha$, $\alpha_{\ell j}$, and $\delta_i$) to avoid high plane wave cut-offs in codes that use periodic boundary conditions.
As a result, some trade-offs in the outcomes arise from these constraints, that prioritize broader applicability overall.
The constructed ccECPs achieve near-optimal accuracy within this given framework. 
\\

Furthermore, we assess transferability by calculating the binding curves $D(r)$ of diatomic molecules at the CCSD(T) level across a broad range of bond lengths, from equilibrium to compressed bond lengths near the dissociation point due to ion-ion repulsion that qualitatively correspond to high pressures in condensed systems, as well as stretched bond lengths approaching the dissociation limit. These energies are compared to fully correlated AE results to measure the quality of the ECPs, with discrepancies calculated as give by
\begin{equation}
    \Delta(r) = D^{\text{ECP}}(r) - D^{\text{AE}}(r)
\end{equation}
The binding curves are then fitted to a Morse potential, extracting parameters such as dissociation energy ($D_e$), equilibrium bond length ($r_e$), and the vibrational parameter ($a$) related to the vibrational frequency of the potential ($\omega_e$), as outlined by:
\begin{equation}
    V(r) = D_e \left(e^{-2a(r - r_e)} - 2e^{-a(r - r_e)} \right),
\end{equation}

\begin{equation}
    \omega_e = \sqrt{\frac{2a^2 D_e}{\mu}},
\end{equation}

\subsection{Selected $5s$ elements}
In this section, we present results for rubidium (Rb) and strontium (Sr), both using a 28-electron core [[Ar]4$d^{10}$] with $s$, $p$, and $d$ non-local channels, consistent with our previous ccECP work\cite{Bennett2018}. For Sr, we developed two variants of ccECP: small-core [[Ar]4$d^{10}$] and large-core [Kr].
\label{Selected 5s elements}
\begin{table*}
\small
\centering
\caption{
SOREP optimized parameters for the selected $5s$ elements ccECPs, with the optimized terms provided as defined in Eqs.(\ref{eq:AREP_local}-\ref{eq:SO}).
$Z_{\text{eff}}$ represents the effective core charge, and $n_{\ell k}$ denotes the Gaussian function form of the channels.
The local channel $L$ corresponds to the largest $\ell$ value, while the highest non-local angular momentum channel is $\ell_{max}=L-1$.
The parameters $\alpha_{\ell k}$ and $\beta_{\ell k}$ are the Gaussian exponents and coefficients, respectively.}
\label{tab:selected_5s_params}
\begin{tabular}{cccccrrccccccrrr}
\hline\hline
\multicolumn{1}{c}{Atom} & \multicolumn{1}{c}{$Z_{\rm eff}$} & \multicolumn{1}{c}{Hamiltonian} & \multicolumn{1}{c}{$\ell$} & \multicolumn{1}{c}{$n_{\ell k}$} & \multicolumn{1}{c}{$\alpha_{\ell k}$} & \multicolumn{1}{c}{$\beta_{\ell k}$} & & \multicolumn{1}{c}{Atom} & \multicolumn{1}{c}{$Z_{\rm eff}$} & \multicolumn{1}{c}{Hamiltonian} & \multicolumn{1}{c}{$\ell$} & \multicolumn{1}{c}{$n_{\ell k}$} & \multicolumn{1}{c}{$\alpha_{\ell k}$} & \multicolumn{1}{c}{$\beta_{\ell k}$} \\
\hline
Rb &  9 & AREP &  0 & 2 &    5.462653 &   89.500251  &   89.500251  && Sr & 10 & AREP &  0 & 2 &    7.439649 &  131.279952  \\
   &    &      &  0 & 2 &    2.056984 &    0.319840  &    0.319839  &&    &    &      &  0 & 2 &    3.447122 &   19.222463  \\
   &    &      &  1 & 2 &    4.588244 &   48.545979  &   48.545978  &&    &    &      &  1 & 2 &    4.744598 &   77.982723  \\
   &    &      &  1 & 2 &    1.032037 &    0.790029  &    0.790029  &&    &    &      &  1 & 2 &    8.237021 &   15.520051  \\
   &    &      &  2 & 2 &    3.413959 &   26.206712  &   26.206712  &&    &    &      &  2 & 2 &    3.183727 &   29.113266  \\
   &    &      &  2 & 2 &    1.024061 &    0.674894  &    0.674894  &&    &    &      &  2 & 2 &    7.748102 &    5.427551  \\
   &    &      &  3 & 1 &    3.500301 &    9.000000  &    9.000000  &&    &    &      &  3 & 1 &    9.273368 &   10.000000  \\
   &    &      &  3 & 3 &    4.112875 &   31.502713  &   31.502713  &&    &    &      &  3 & 3 &    9.163478 &   92.733680  \\
   &    &      &  3 & 2 &    3.366937 &  -10.463215  &  -10.463214  &&    &    &      &  3 & 2 &   10.340438 &   -4.567422  \\
   &    &      &  3 & 2 &    1.707218 &   -2.167674  &   -2.167673  &&    &    &      &  3 & 2 &    9.152969 &  -95.475410  \\
   &    &      &    &   &             &              &              &&    &    &      &    &   &             &              \\
   &    & SO   &  1 & 2 &    4.388251 &  -39.040315  &  -39.040315  &&    &    & SO   &  1 & 2 &    7.222917 &  -58.876406  \\
   &    &      &  1 & 2 &    4.327577 &   39.039369  &   39.039369  &&    &    &      &  1 & 2 &    7.171633 &   58.880316  \\
   &    &      &  1 & 2 &    1.357223 &   -0.331685  &   -0.331685  &&    &    &      &  1 & 2 &    3.003530 &   -9.875334  \\
   &    &      &  1 & 2 &    1.230710 &    0.437135  &    0.437134  &&    &    &      &  1 & 2 &    2.833933 &    9.735187  \\
   &    &      &  2 & 2 &    3.409706 &  -10.483339  &  -10.483338  &&    &    &      &  2 & 2 &    6.319959 &  -11.907111  \\
   &    &      &  2 & 2 &    3.418975 &   10.482132  &   10.482131  &&    &    &      &  2 & 2 &    6.391156 &   11.906538  \\
   &    &      &  2 & 2 &    1.022484 &   -0.279420  &   -0.279419  &&    &    &      &  2 & 2 &    1.767711 &   -2.201784  \\
   &    &      &  2 & 2 &    1.027724 &    0.260697  &    0.260697  &&    &    &      &  2 & 2 &    1.637060 &    1.928907  \\
   &    &      &    &   &             &              &              &&    &    &      &    &   &             &              \\
   &    &      &    &   &             &              &              && Sr &  2 & AREP &  0 & 2 &    0.825002 &   15.386176  \\
   &    &      &    &   &             &              &              &&    &    &      &  1 & 2 &    0.397788 &    5.077782  \\
   &    &      &    &   &             &              &              &&    &    &      &  2 & 2 &    0.453259 &   -2.256103  \\
   &    &      &    &   &             &              &              &&    &    &      &  3 & 1 &    3.000056 &    2.000000  \\
   &    &      &    &   &             &              &              &&    &    &      &  3 & 3 &    3.002302 &    6.000111  \\
   &    &      &    &   &             &              &              &&    &    &      &  3 & 2 &    2.996531 &  -10.000277  \\
   &    &      &    &   &             &              &              &&    &    &      &    &   &             &              \\
\hline\hline
\end{tabular}
\end{table*}
\subsubsection{Rb}
\begin{figure*}[!htbp]
\centering
\begin{subfigure}{0.5\textwidth}
\includegraphics[width=\textwidth]{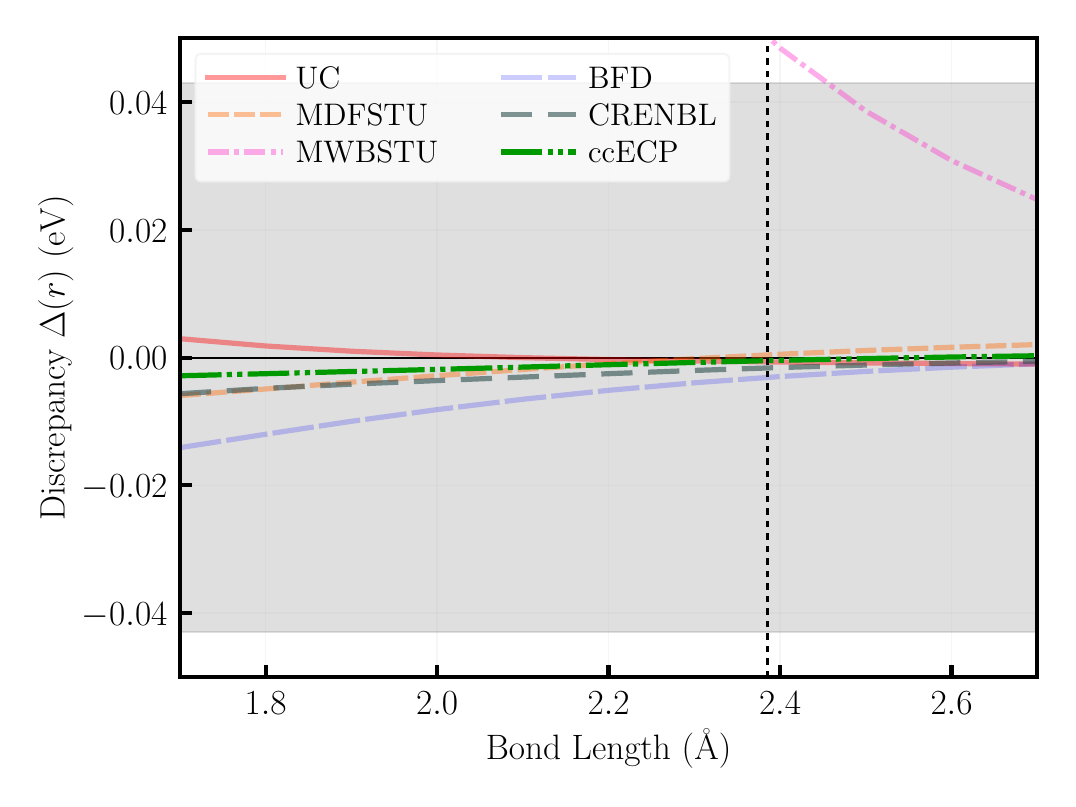}
\caption{RbH binding curve discrepancies}
\label{fig:RbH}
\end{subfigure}%
\begin{subfigure}{0.5\textwidth}
\includegraphics[width=\textwidth]{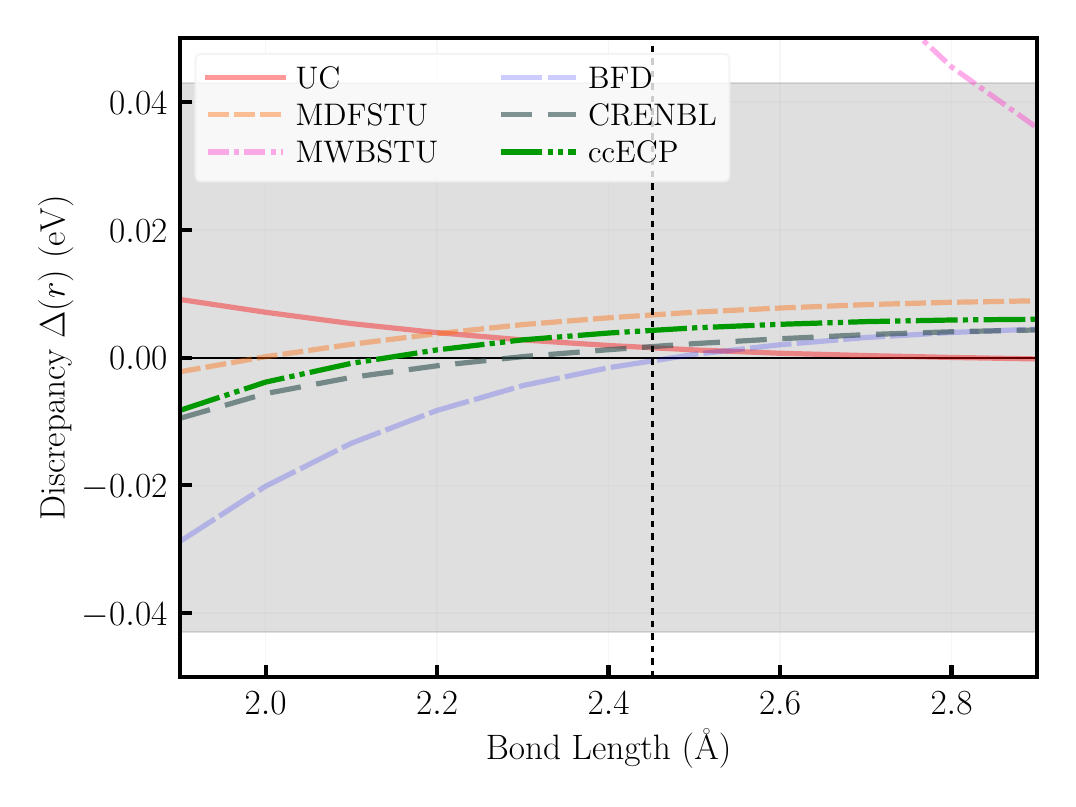}
\caption{RbO binding curve discrepancies}
\label{fig:RbO}
\end{subfigure}
\caption{Binding energy discrepancies for (a) RbH and (b) RbO molecules. The gray-shaded area indicates the region of chemical accuracy, and the vertical dashed line indicates the equilibrium bond length from the Morse potential fit of the AE data. We also tested existing legacy ECPs using core definitions consistent with the ccECP.}
\label{fig:Rb_mols}
\end{figure*}

Rb has closed $4s$ and $4p$ shells and a half-filled $5s$ shell. This configuration with 9 valence electrons lends itself to developing a robustly performing ECP. From Figs. \ref{fig:MAD_in_elements} and \ref{fig:Rb_mols} we see that the legacy ECPs and our ccECP perform well within chemical accuracy in the atomic and molecular spectra. Only, LANL2 deviates significantly from the AE atomic spectrum and this discrepancy is amplified in the molecular properties. Therefore, in order to provide a better comparison between other ECPs, LANL2 has been omitted from the molecular binding curves as it was out of bounds. For the [[Ar]3$d^{10}$] core, there are few ECPs available to compare against, and our ccECP shows a slight overall improvement specifically close to the equilibrium bond lengths and in the LMAD criterion.
\subsubsection{Sr}
\begin{figure*}[!htbp]
\centering
\begin{subfigure}{0.5\textwidth}
\includegraphics[width=\textwidth]{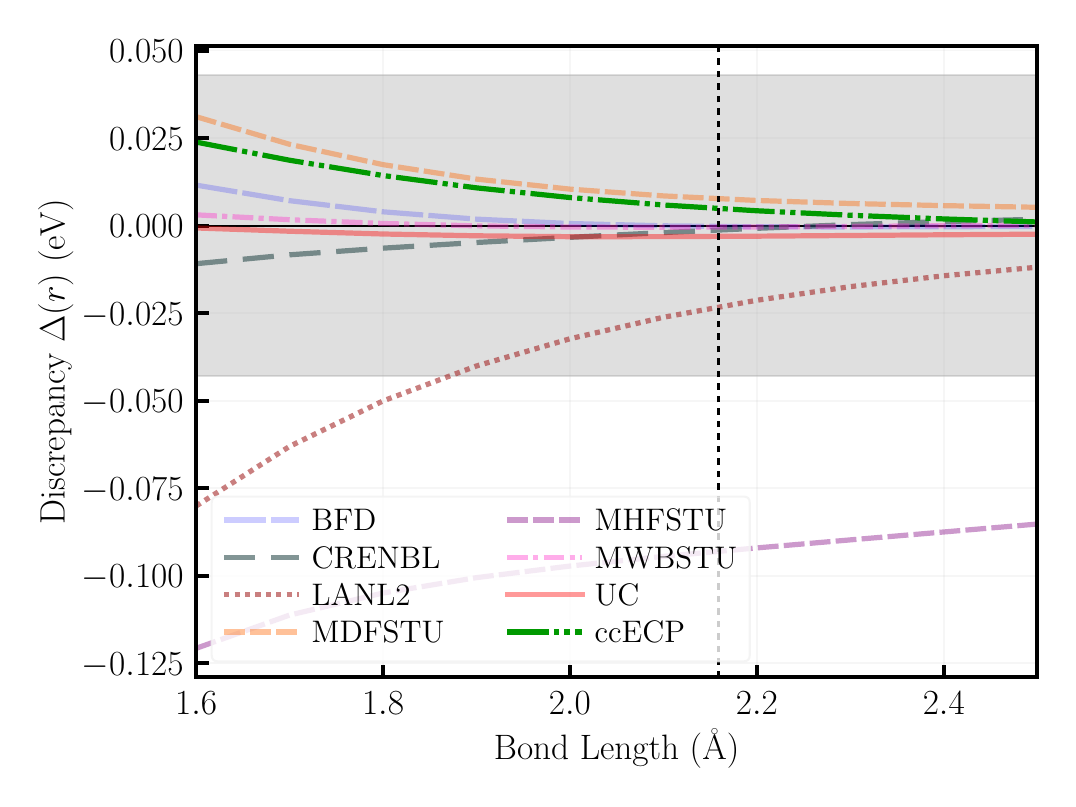}
\caption{SrH binding curve discrepancies}
\label{fig:SrH28}
\end{subfigure}%
\begin{subfigure}{0.5\textwidth}
\includegraphics[width=\textwidth]{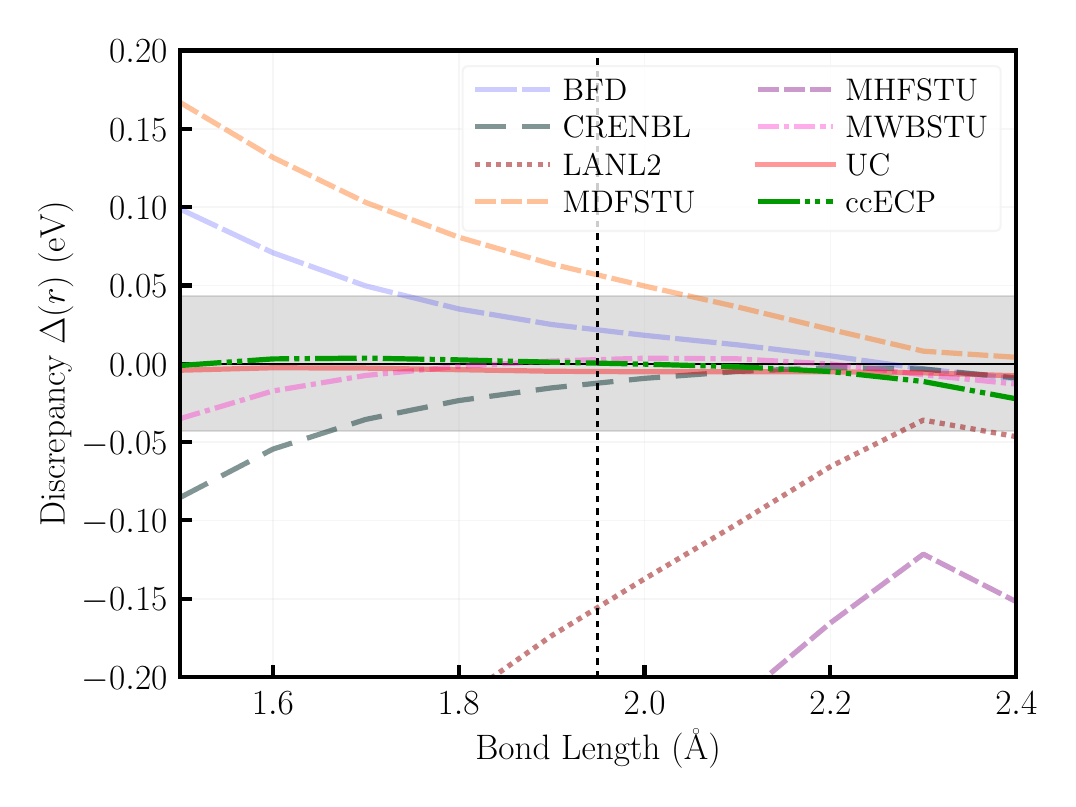}
\caption{SrO binding curve discrepancies}
\label{fig:SrO28}
\end{subfigure}
\caption{Binding energy discrepancies for (a) SrH and (b) SrO molecules using the small-core [[Ar]3$d^{10}$]. The same conventions apply as in Fig. \ref{fig:Rb_mols}.}
\label{fig:Sr28_mols}
\end{figure*}

\begin{figure*}[!htbp]
\centering
\begin{subfigure}{0.5\textwidth}
\includegraphics[width=\textwidth]{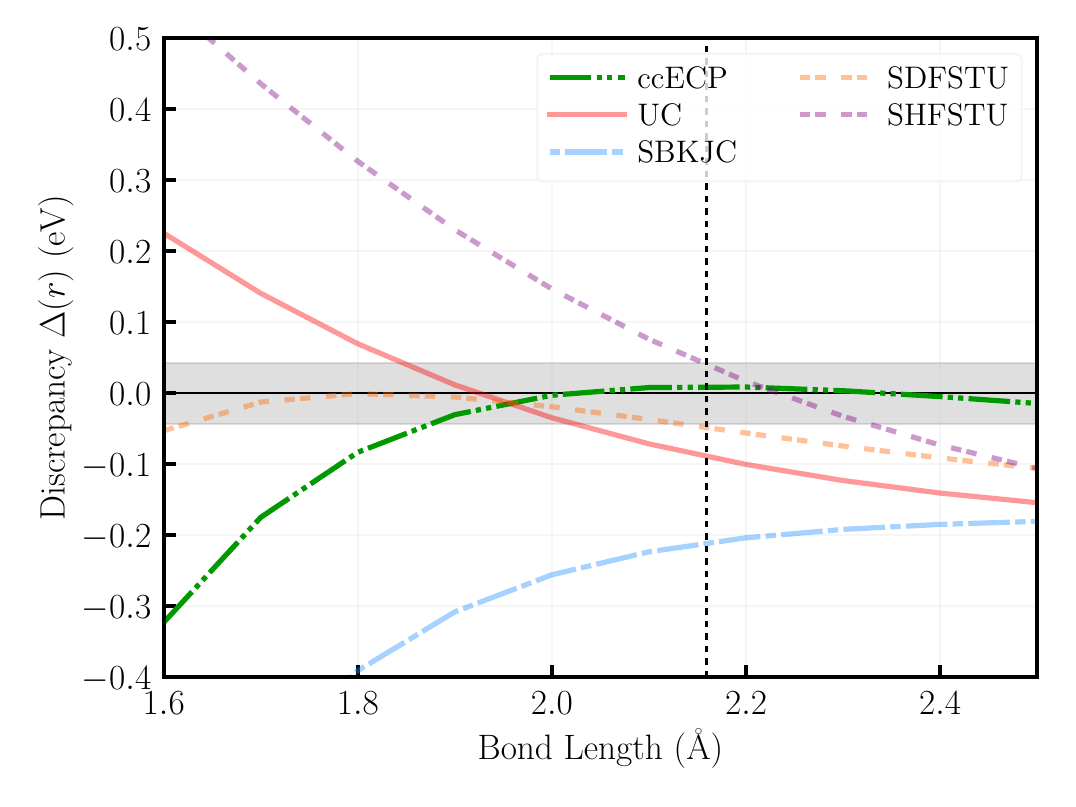}
\caption{SrH binding curve discrepancies}
\label{fig:SrH36}
\end{subfigure}%
\begin{subfigure}{0.5\textwidth}
\includegraphics[width=\textwidth]{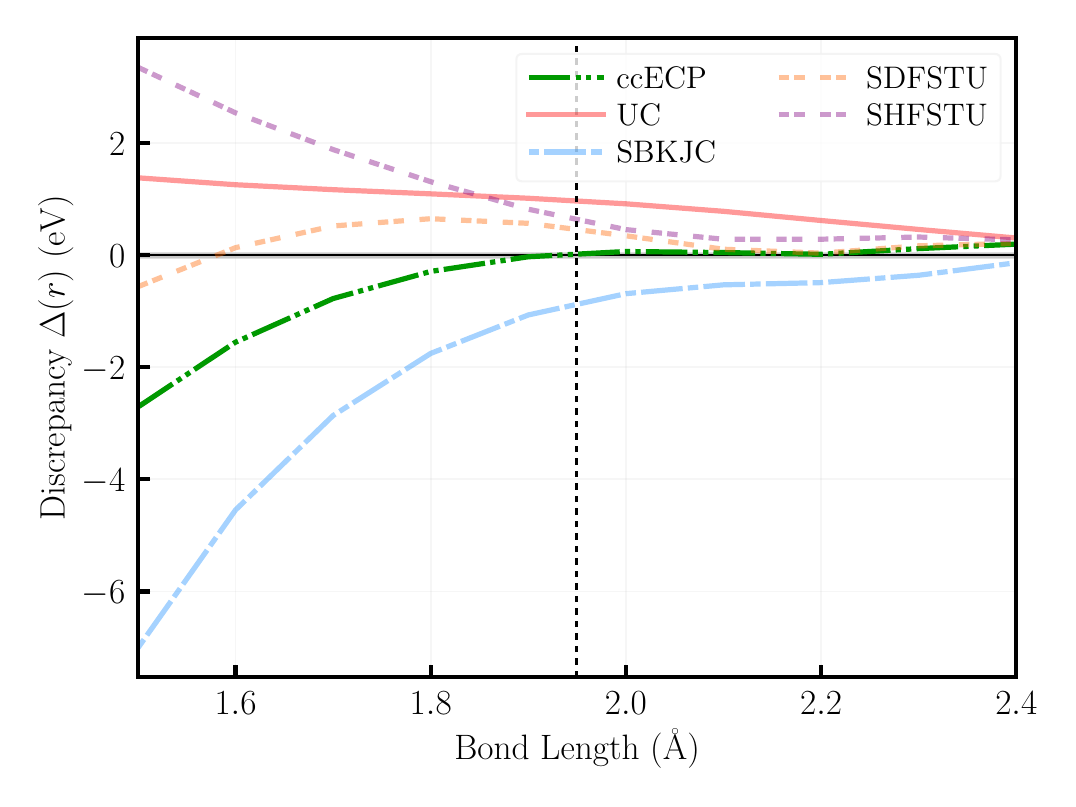}
\caption{SrO binding curve discrepancies}
\label{fig:SrO36}
\end{subfigure}
\caption{Binding energy discrepancies for (a) SrH and (b) SrO molecules  using the large-core [Kr]. Conventions are the same as defined previously in Fig. \ref{fig:Rb_mols}.}
\label{fig:Sr36_mols}
\end{figure*}

For Sr, the small-core [[Ar]3$d^{10}$] ccECP performs well in the atomic spectrum, matching the accuracy of CRENBL and UC. For SrH hydride (Fig. \ref{fig:Sr28_mols}), which has a relatively simple electronic structure, the ccECP achieves a chemical accuracy comparable to that of legacy ECPs. For SrO, which is a more challenging system, the ccECP shows minimal discrepancies ($\lesssim$ 0.02 ev) across most bond lengths and matches the performance of UC. 
Compared to the small-core, the large-core [Kr] 
variant shows noticeable differences in isospectrality (Figs. \ref{fig:MAD_in_elements}-\ref{fig:WMAD_in_elements}) and in molecular binding curves (Fig. \ref{fig:Sr36_mols}). In the atomic spectrum, the ccECP outperforms all legacy ECPs and the UC results, with MAD values below 0.1 eV. In molecules, despite the challenge of having only two valence electrons, our ccECP is the only ECP that perform consistently within the bounds of chemical accuracy across many of the geometries tested. The ccECP offers negligible discrepancies at the equilibrium bond length, approaching the limits of this construction. Although the large core variant is efficient, the small core [[Ar]4$d^{10}$] remains viable for studies that require a proper description of semi-core levels. 
\subsection{Selected $6s$ elements}
We constructed the ccECPs for cesium (Cs) and barium (Ba) using a [[Kr]4$d^{10}]$] core while including $5s^25p^2$ into the valence space. We could not include SBKJC in our tests as it lacks a [[Kr]4$d^{10}$] core ECP.
\begin{table*}
\small
\centering
\caption{SOREP optimized parameters for the selected $6s$ elements ccECPs. The parameters follow the same definitions as in Table \ref{tab:selected_5s_params}.
}
\label{tab:selected_6s_params}
\begin{tabular}{cccccrrccccccrrr}
\hline\hline
\multicolumn{1}{c}{Atom} & \multicolumn{1}{c}{$Z_{\rm eff}$} & \multicolumn{1}{c}{Hamiltonian} & \multicolumn{1}{c}{$\ell$} & \multicolumn{1}{c}{$n_{\ell k}$} & \multicolumn{1}{c}{$\alpha_{\ell k}$} & \multicolumn{1}{c}{$\beta_{\ell k}$} & & \multicolumn{1}{c}{Atom} & \multicolumn{1}{c}{$Z_{\rm eff}$} & \multicolumn{1}{c}{Hamiltonian} & \multicolumn{1}{c}{$\ell$} & \multicolumn{1}{c}{$n_{\ell k}$} & \multicolumn{1}{c}{$\alpha_{\ell k}$} & \multicolumn{1}{c}{$\beta_{\ell k}$} \\
\hline
Cs & 9  & AREP &  0 & 2 &    4.077065 &   85.047793  && Ba & 10 & AREP &  0 & 2 &    4.046183 &   84.785943 \\
   &    &      &  0 & 2 &    2.417524 &   16.654293  &&    &    &      &  0 & 2 &    2.258112 &   17.375717 \\
   &    &      &  1 & 2 &    5.521131 &  156.549042  &&    &    &      &  1 & 2 &    6.361827 &  157.532338 \\
   &    &      &  1 & 2 &    2.202089 &   26.422928  &&    &    &      &  1 & 2 &    2.002936 &   25.777693 \\
   &    &      &  2 & 2 &    2.010196 &   13.672640  &&    &    &      &  2 & 2 &    1.863849 &   12.467708 \\
   &    &      &  2 & 2 &    0.865141 &    3.342155  &&    &    &      &  2 & 2 &    0.848671 &    3.415276 \\
   &    &      &  3 & 1 &    2.999986 &    9.000000  &&    &    &      &  3 & 2 &    6.200000 &  -46.100000 \\
   &    &      &  3 & 3 &    2.999542 &   26.999875  &&    &    &      &  3 & 2 &    1.620000 &   -5.320000 \\
   &    &      &  3 & 2 &    3.000404 &   -9.999908  &&    &    &      &  4 & 1 &    9.999991 &   10.000000 \\
   &    &      &  3 & 2 &    3.324776 &  -34.269488  &&    &    &      &  4 & 3 &    9.998491 &   99.999090 \\
   &    &      &    &   &             &              &&    &    &      &  4 & 2 &    9.997459 &  120.999962 \\
   &    &      &    &   &             &              &&    &    &      &  4 & 2 &    2.263468 &   -7.579943 \\
   &    &      &    &   &             &              &&    &    &      &    &   &             &             \\
   &    & SO   &  1 & 2 &    7.254842 &  -58.875111  &&    &    &  SO  &  1 & 2 &    8.049117 & -104.977838 \\
   &    &      &  1 & 2 &    7.134591 &   58.881762  &&    &    &      &  1 & 2 &    5.815014 &  105.053228 \\
   &    &      &  1 & 2 &    3.350131 &   -9.842037  &&    &    &      &  1 & 2 &    1.699651 &  -17.450004 \\
   &    &      &  1 & 2 &    2.500019 &    9.757508  &&    &    &      &  1 & 2 &    1.684836 &   17.205650 \\
   &    &      &  2 & 2 &    6.322741 &  -11.906934  &&    &    &      &  2 & 2 &    1.021293 &   -4.864392 \\
   &    &      &  2 & 2 &    6.390190 &   11.906574  &&    &    &      &  2 & 2 &    0.972565 &    5.606032 \\
   &    &      &  2 & 2 &    1.797091 &   -2.190492  &&    &    &      &  2 & 2 &    0.523179 &   -1.093392 \\
   &    &      &  2 & 2 &    1.603969 &    1.940152  &&    &    &      &  2 & 2 &    0.305141 &    0.334212 \\
   &    &      &    &   &             &              &&    &    &      &  3 & 2 &    4.488933 &   15.335482 \\
   &    &      &    &   &             &              &&    &    &      &  3 & 2 &    8.134136 &  -11.059107 \\
   &    &      &    &   &             &              &&    &    &      &  3 & 2 &    3.339176 &    3.562972 \\
   &    &      &    &   &             &              &&    &    &      &  3 & 2 &    0.811040 &   -1.372854 \\
   &    &      &    &   &             &              &&    &    &      &    &   &            &              \\
\hline\hline
\end{tabular}
\end{table*}
\subsubsection{Cs}
\begin{figure*}[!htbp]
\centering
\begin{subfigure}{0.5\textwidth}
\includegraphics[width=\textwidth]{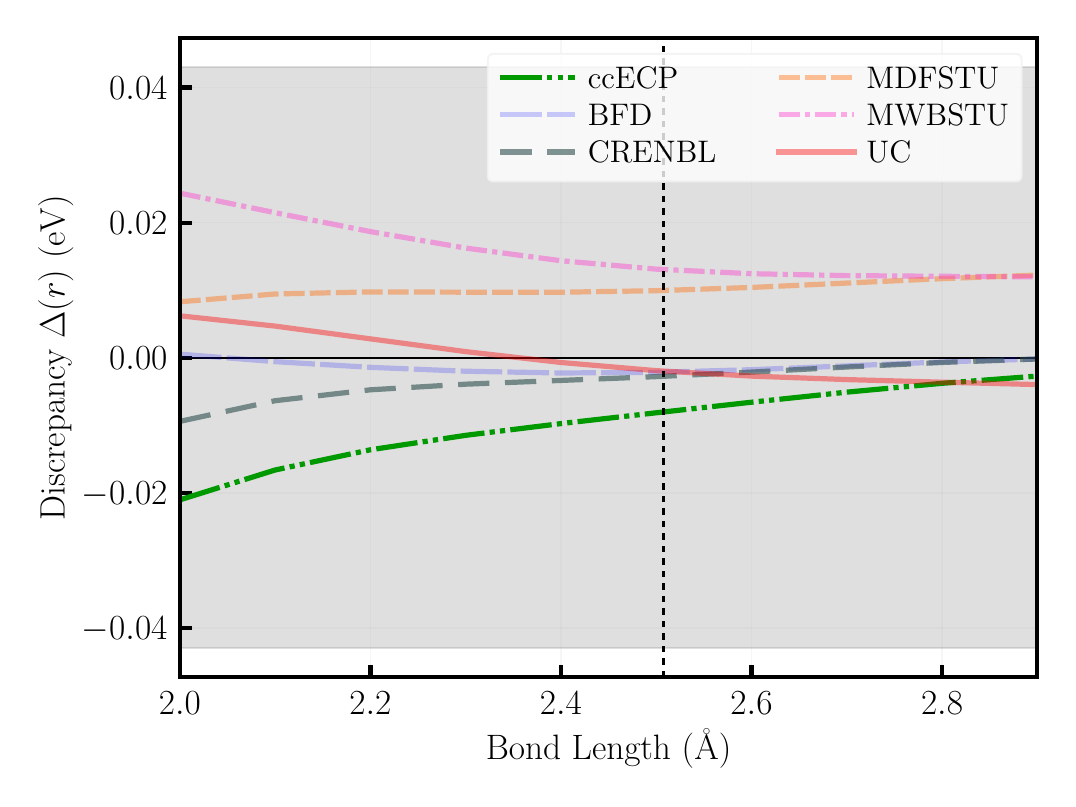}
\caption{CsH binding curve discrepancies}
\label{fig:CsH}
\end{subfigure}%
\begin{subfigure}{0.5\textwidth}
\includegraphics[width=\textwidth]{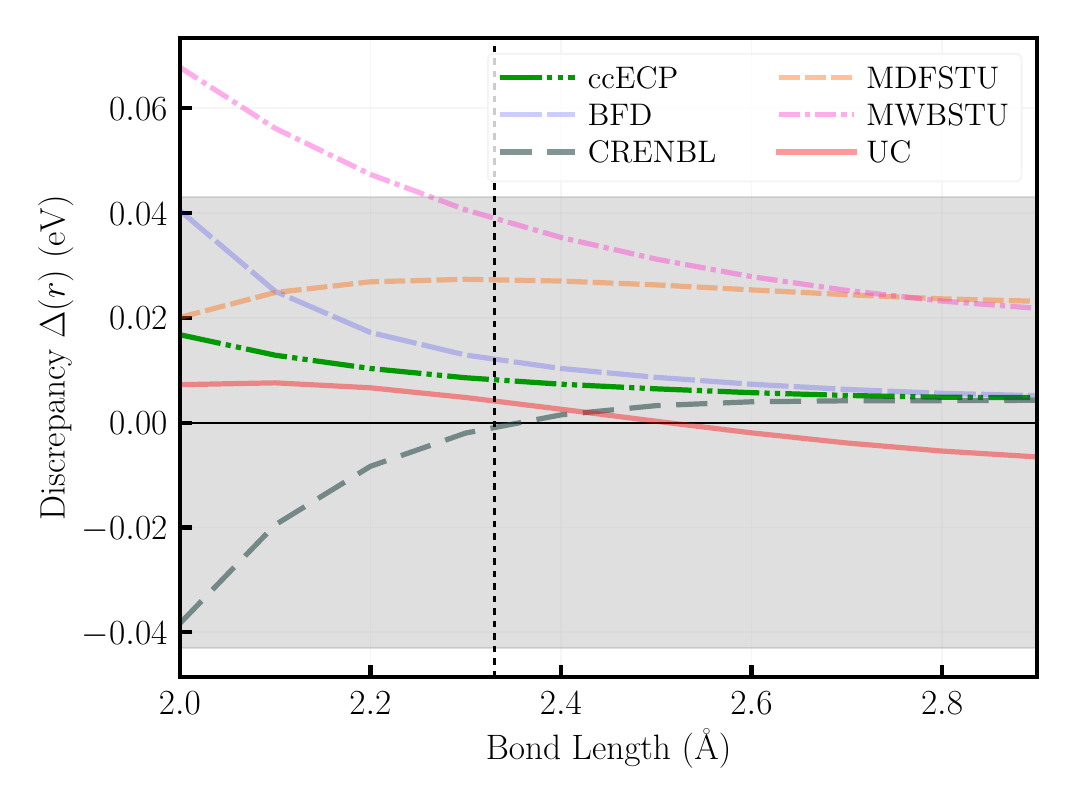}
\caption{CsO binding curve discrepancies}
\label{fig:CsO}
\end{subfigure}
\caption{Binding energy discrepancies for (a) CsH and (b) CsO molecules. LANL2 was tested as well, but the results lie outside of the region graphed. Conventions are the same as defined previously in Fig. \ref{fig:Rb_mols}. MHFSTU is not plotted as it lays far outside of the region shown.}
\label{fig:Cs_mols}
\end{figure*}

The simple electronic structure of Cs bonding is reflected in the CsH and CsO molecules in Fig. \ref{fig:Cs_mols} where most of the ECPs are well within the chemical accuracy regime. The ccECP shows noticeable improvements at the equilibrium bond length (Fig. \ref{fig:Cs_mols}). Similarly to the case of Rb, LANL2 has been omitted from the molecular binding curves due to its large discrepancies. Furthermore, for the ccECP we observe a systematic improvement in the atomic spectrum compared to the legacy ECPs, as shown in Figs. \ref{fig:MAD_in_elements}-\ref{fig:WMAD_in_elements} 
\subsubsection{Ba}
\begin{figure*}[!htbp]
\centering
\begin{subfigure}{0.5\textwidth}
\includegraphics[width=\textwidth]{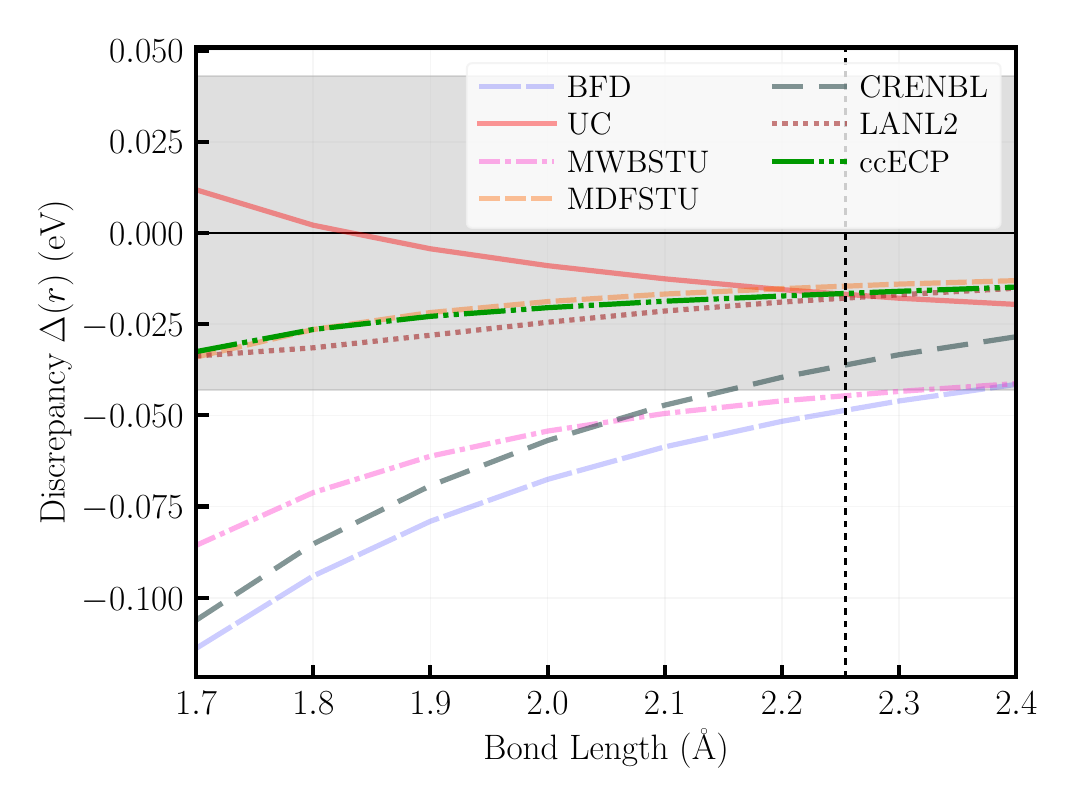}
\caption{BaH binding curve discrepancies}
\label{fig:BaH}
\end{subfigure}%
\begin{subfigure}{0.5\textwidth}
\includegraphics[width=\textwidth]{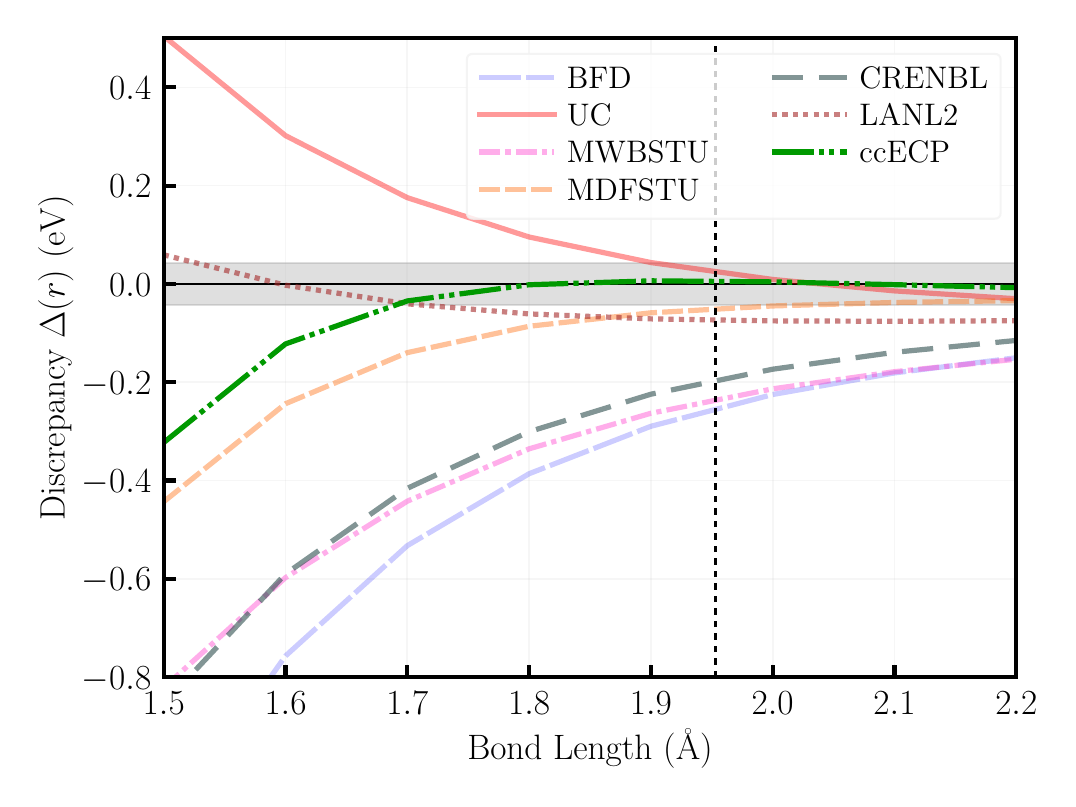}
\caption{BaO binding curve discrepancies}
\label{fig:BaO}
\end{subfigure}
\caption{Binding energy discrepancies for (a) BaH and (b) BaO molecules. Conventions are the same as defined previously in Fig. \ref{fig:Rb_mols}.}
\label{fig:Ba_mols}
\end{figure*}

Although Ba has no occupied $f$ electrons in its ground state, the $f$ channel is crucial to capture subtle electronic interactions and relativistic effects, ensuring a more precise representation of this atom in complex systems. During optimization, we sought a balance between accurately replicating the AE atomic spectrum and maintaining robust performance across various chemical environments. Our primary focus was to prioritize precision in the atomic spectrum, accepting a minor trade-off in molecular binding curves (Fig. \ref{fig:BaO}) for BaO at compressed bond lengths. The ccECP demonstrates consistent results within chemical accuracy across a wide array of bond lengths, particularly at the equilibrium geometry. During the optimization step we focused on mitigating the overbinding issue at compressed bond lengths, achieving higher accuracy compared to several legacy ECPs and notably outperforming even the UC method in this critical region. In our view, this approach represents the optimal balance, maximizing the overall efficacy and reliability of the ccECP. 
\subsection{Selected $5p$ and $6p$ elements}
For $5p$ and $6p$ main group elements we focused on [[Kr]4$d^{10}$] and [[Xe]4$f^{14}$5$d^{10}$] cores, respectively, as these are the most commonly chosen partitionings for these elements. However, in the case of indium (In), we explored an alternate core with [[Ar]3$d^{10}$], following reasoning similar to that outlined for Sr in Sec. \ref{Selected 5s elements}. 

\begin{table*}
\small
\centering
\caption{SOREP optimized parameters for the selected $5p$ and $6p$ elements ccECPs. The parameters follow the same definitions as in Table \ref{tab:selected_5s_params}.
}
\label{tab:selected_5p_and_6p_params}
\begin{tabular}{cccccrrccccccrrr}
\hline\hline
\multicolumn{1}{c}{Atom} & \multicolumn{1}{c}{$Z_{\rm eff}$} & \multicolumn{1}{c}{Hamiltonian} & \multicolumn{1}{c}{$\ell$} & \multicolumn{1}{c}{$n_{\ell k}$} & \multicolumn{1}{c}{$\alpha_{\ell k}$} & \multicolumn{1}{c}{$\beta_{\ell k}$} & & \multicolumn{1}{c}{Atom} & \multicolumn{1}{c}{$Z_{\rm eff}$} & \multicolumn{1}{c}{Hamiltonian} & \multicolumn{1}{c}{$\ell$} & \multicolumn{1}{c}{$n_{\ell k}$} & \multicolumn{1}{c}{$\alpha_{\ell k}$} & \multicolumn{1}{c}{$\beta_{\ell k}$} \\
\hline
In & 21 & AREP &  0 & 2 &   13.668243 &  270.034991  && Sb &  5 & AREP &  0 & 2 &    2.332041 &   68.292881  \\
   &    &      &  0 & 2 &    7.552829 &   38.853313  &&    &    &      &  0 & 2 &    1.376531 &   -7.220586  \\
   &    &      &  1 & 2 &   12.664151 &  193.825370  &&    &    &      &  1 & 2 &    2.161816 &   54.497670  \\
   &    &      &  1 & 2 &    6.802069 &   31.859027  &&    &    &      &  1 & 2 &    1.219158 &   -1.963704  \\
   &    &      &  2 & 2 &   10.614660 &   79.192746  &&    &    &      &  2 & 2 &    0.930013 &    3.774218  \\
   &    &      &  2 & 2 &    4.898782 &   12.851614  &&    &    &      &  2 & 2 &    0.912651 &    5.637552  \\
   &    &      &  3 & 1 &   10.993168 &   21.000000  &&    &    &      &  3 & 1 &    1.500000 &    5.000000  \\
   &    &      &  3 & 3 &   10.712646 &  230.856524  &&    &    &      &  3 & 3 &    1.500000 &    7.500000  \\
   &    &      &  3 & 2 &   11.560723 & -164.383958  &&    &    &      &  3 & 2 &    1.619327 &  -14.984235  \\
   &    &      &  3 & 2 &   11.750247 &  -29.364920  &&    &    &      &  3 & 2 &    1.500000 &   -1.950880  \\
   &    &      &    &   &             &              &&    &    &      &    &   &             &  \\
   &    & SO   &  1 & 2 &   13.906173 & -134.925215  &&    &    & SO   &  1 & 2 &    2.135386 &  -37.733817  \\
   &    &      &  1 & 2 &   13.373690 &  134.948225  &&    &    &      &  1 & 2 &    2.136905 &   37.717372  \\
   &    &      &  1 & 2 &    7.585881 &  -29.495887  &&    &    &      &  1 & 2 &    1.058461 &    1.488076  \\
   &    &      &  1 & 2 &    7.350840 &   29.635242  &&    &    &      &  1 & 2 &    1.495239 &   -1.498848  \\
   &    &      &  2 & 2 &   14.043454 &  -35.491736  &&    &    &      &  2 & 2 &    0.936191 &   -3.771865  \\
   &    &      &  2 & 2 &   14.503687 &   35.453916  &&    &    &      &  2 & 2 &    0.916391 &    3.757636  \\
   &    &      &  2 & 2 &    5.579398 &   -9.167354  &&    &    &      &    &   &             &              \\
   &    &      &  2 & 2 &    5.027172 &    8.273383  &&    &    &      &    &   &             &              \\
   &    &      &    &   &             &              &&    &    &      &    &   &             &              \\
   &    &      &    &   &             &              &&    &    &      &    &   &             &              \\
In & 3  & AREP &  0 & 2 &    1.415769 &   29.165278  && Pb & 4  & AREP &  0 & 2 &    1.989794 &   35.774361  \\    
   &    &      &  0 & 2 &    0.641501 &   -4.203233  &&    &    &      &  0 & 2 &    0.254518 &   -0.564234  \\
   &    &      &  1 & 2 &    1.468724 &   36.990136  &&    &    &      &  1 & 2 &    0.922154 &    8.163927  \\
   &    &      &  1 & 2 &    0.682829 &   -3.370655  &&    &    &      &  1 & 2 &    0.227206 &   -0.252977  \\
   &    &      &  2 & 2 &    0.782626 &   10.066731  &&    &    &      &  2 & 2 &    0.601501 &    2.960413  \\
   &    &      &  2 & 2 &    0.932383 &    9.934578  &&    &    &      &  2 & 2 &    0.530396 &    4.248686  \\
   &    &      &  3 & 1 &    0.979669 &    3.000000  &&    &    &      &  3 & 2 &    0.813526 &   -2.145000  \\
   &    &      &  3 & 3 &    0.892584 &    2.939008  &&    &    &      &  3 & 2 &    0.778976 &   -2.666419  \\
   &    &      &  3 & 2 &    0.967821 &    0.803666  &&    &    &      &  4 & 1 &    1.998442 &    4.000000  \\
   &    &      &  3 & 2 &    0.839106 &   -6.025764  &&    &    &      &  4 & 3 &    2.007475 &    7.993766  \\
   &    &      &    &   &             &              &&    &    &      &  4 & 2 &    1.988107 &  -12.707015  \\
   &    &      &    &   &             &              &&    &    &      &  4 & 2 &    0.997975 &   -0.235863  \\
   &    &      &    &   &             &              &&    &    &      &    &   &             &              \\
   &    &      &    &   &             &              &&    &    & SO   &  1 & 2 &    1.049108 &   -5.281458  \\
   &    &      &    &   &             &              &&    &    &      &  1 & 2 &    0.985610 &    5.514626  \\
   &    &      &    &   &             &              &&    &    &      &  1 & 2 &    0.271514 &    0.271539  \\
   &    &      &    &   &             &              &&    &    &      &  1 & 2 &    0.190417 &   -0.078010  \\
   &    &      &    &   &             &              &&    &    &      &  2 & 2 &    0.476668 &   -3.426579  \\
   &    &      &    &   &             &              &&    &    &      &  2 & 2 &    0.417629 &    2.573093  \\
   &    &      &    &   &             &              &&    &    &      &  3 & 2 &    0.813760 &    1.429972  \\
   &    &      &    &   &             &              &&    &    &      &  3 & 2 &    0.779307 &   -1.333190  \\   
   &    &      &    &   &             &              &&    &    &      &    &   &             &              \\
\hline\hline
\end{tabular}
\end{table*}

\subsubsection{In}
Out of the two core choices for In, the small-core [[Ar]3$d^{10}$] is designed for smaller systems with fewer electrons where precision is specifically paramount and computational cost is not as much of a concern.
Due to the large number of electrons in the valence space, reproducing the all-electron results is fairly straightforward.
The small-core In ccECP has the best performance in the LMADs, with even smaller average discrepancies than the UC method, while remaining competitive in the other metrics. Within the molecules, all ECPs tested, including our ccECP, are within chemical accuracy for all bond lengths.

\begin{figure*}[!htbp]
\centering
\begin{subfigure}{0.5\textwidth}
\includegraphics[width=\textwidth]{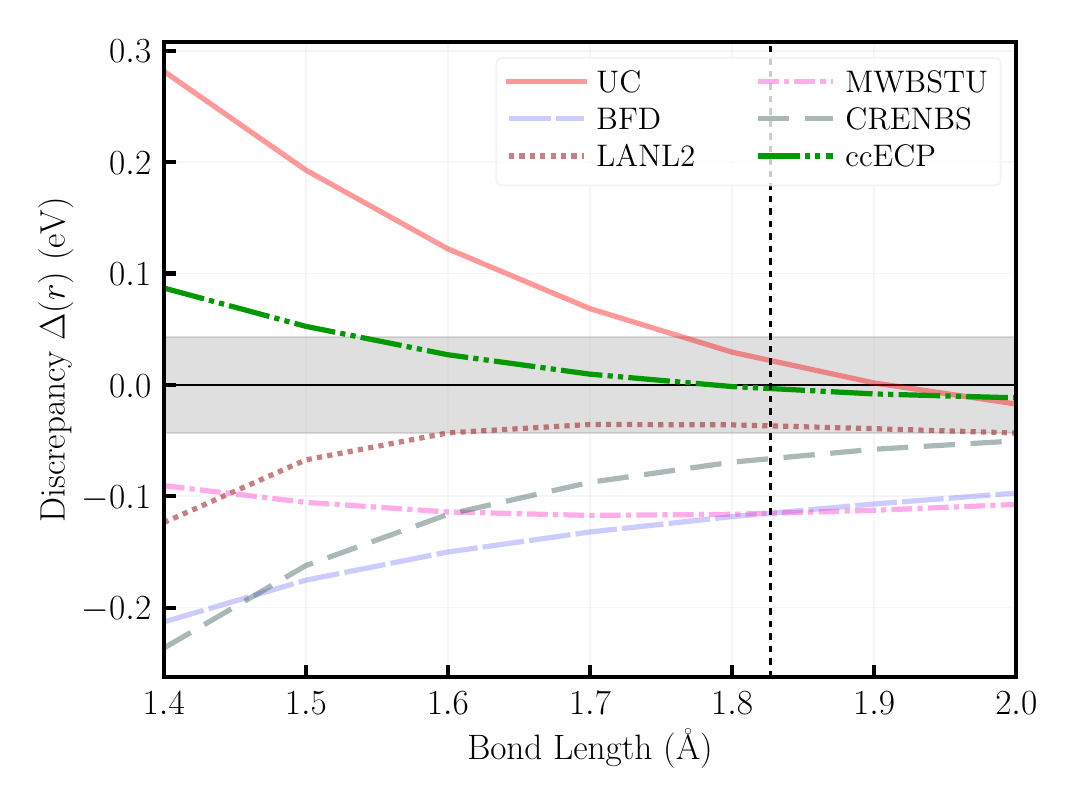}
\caption{InH binding curve discrepancies}
\label{fig:InH46}
\end{subfigure}%
\begin{subfigure}{0.5\textwidth}
\includegraphics[width=\textwidth]{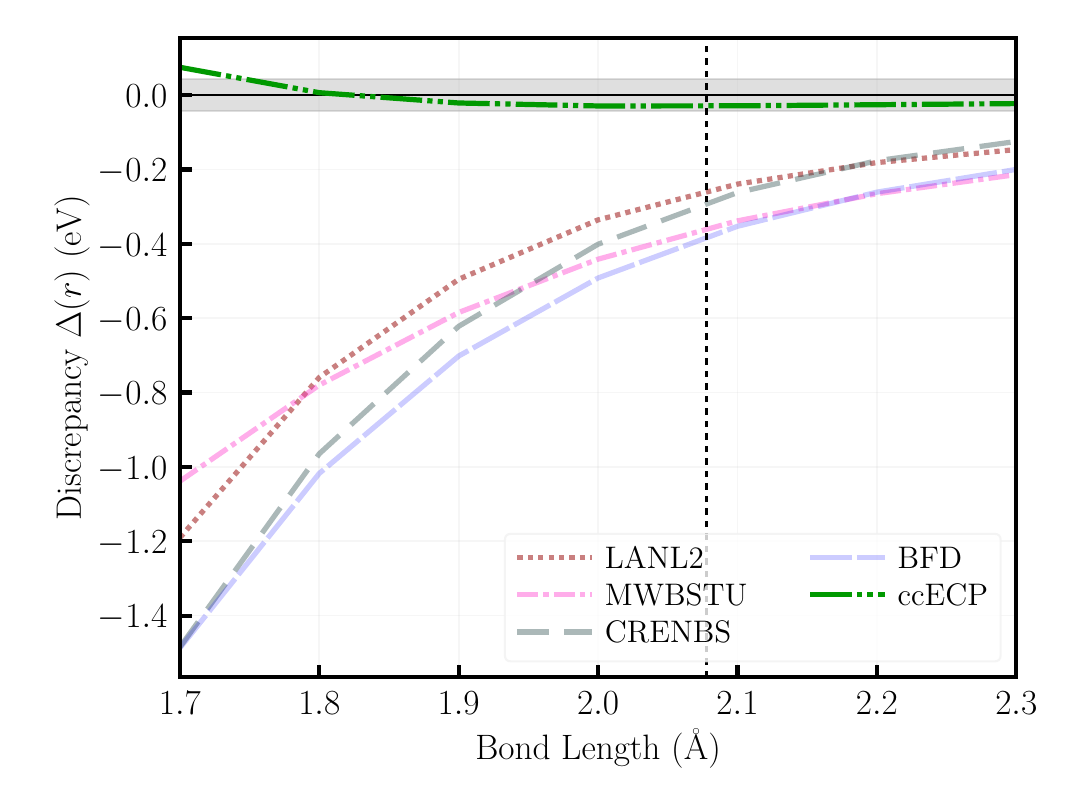}
\caption{InO binding curve discrepancies}
\label{fig:InO46}
\end{subfigure}
\caption{Binding energy discrepancies for (a) InH and (b) InO molecules using large-core [[Kr]4$d^{10}$].
Conventions are the same as defined previously in Fig. \ref{fig:Rb_mols}.}
\end{figure*}

Coming to the large core In [[Kr]4$d^{10}$], with only 3 electrons in the valence space, it can be seen that despite the limited number of valence electrons, it outperforms all other legacy ECPs and even UC in the atomic spectrum. In addition, for the case of molecules, it is the only ECP consistently within the chemical accuracy around the equilibirium bond length for both InH and InO. \ref{fig:InO46}. 
We note that since the valence space is very small, only a limited number of states are able to be meaningfully tested in the atom.
Also for the large core version, we do not present UC results for InO as we were unable to converge it to the proper state. 

\begin{figure*}[!htbp]
\centering
\begin{subfigure}{0.5\textwidth}
\includegraphics[width=\textwidth]{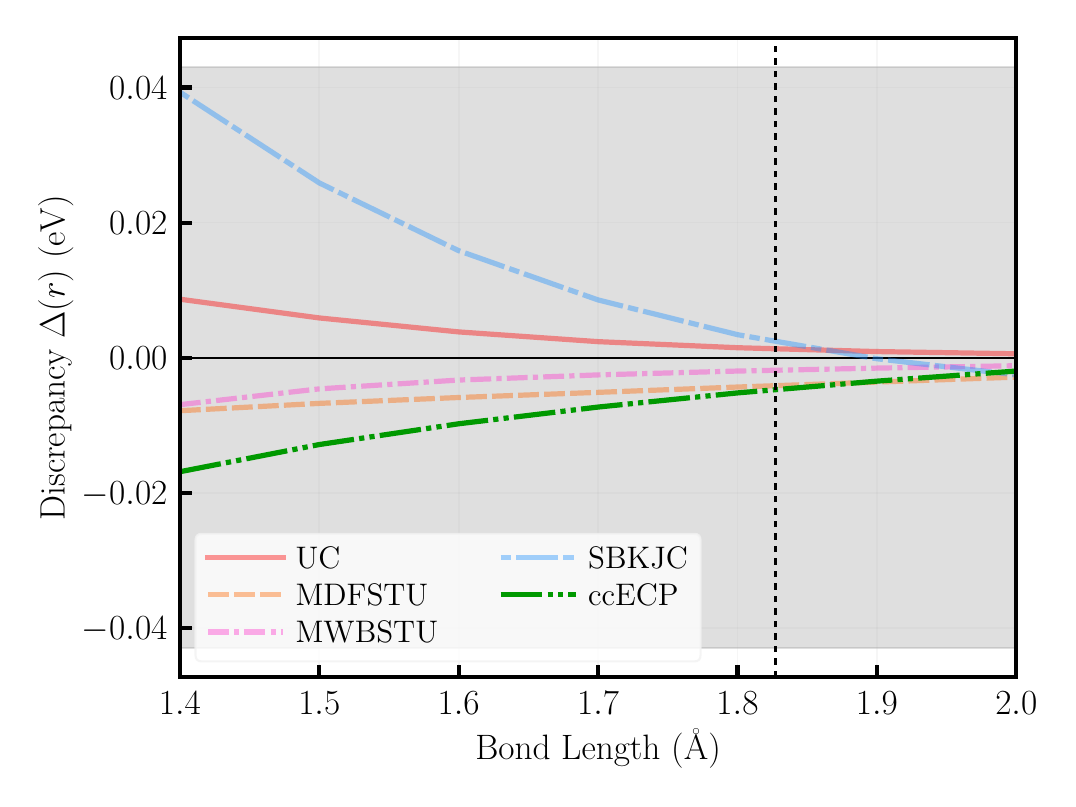}
\caption{InH binding curve discrepancies}
\label{fig:InH28}
\end{subfigure}%
\begin{subfigure}{0.5\textwidth}
\includegraphics[width=\textwidth]{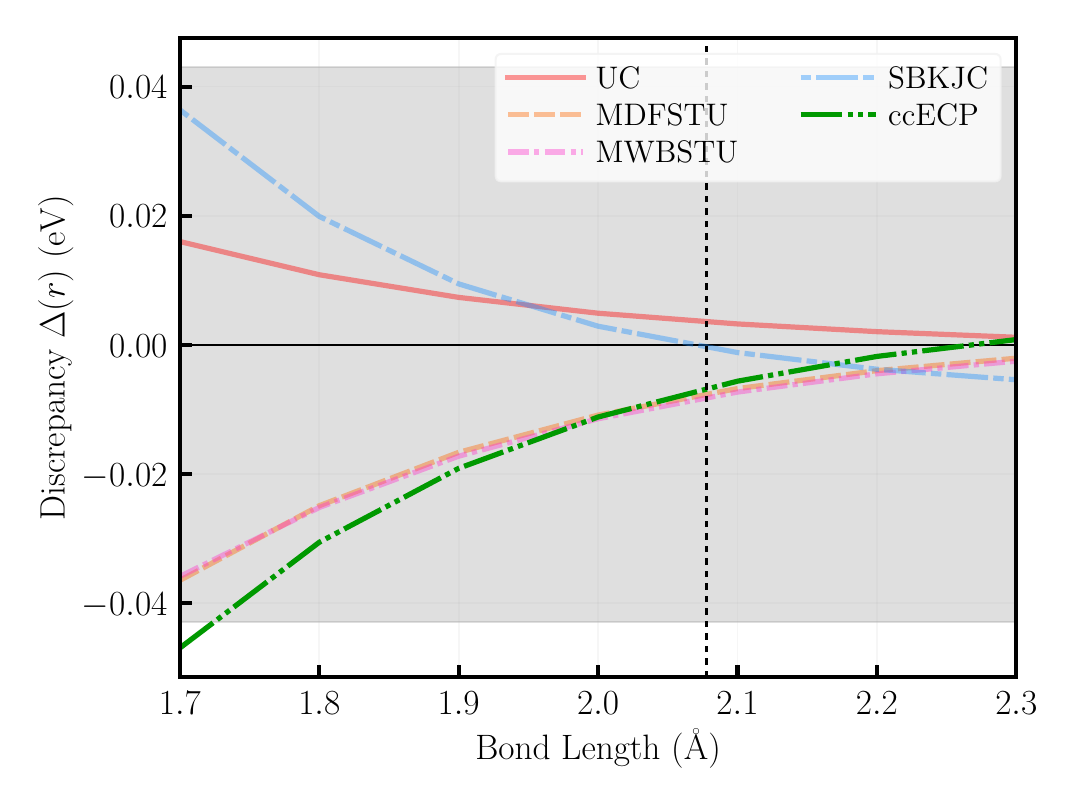}
\caption{InO binding curve discrepancies}
\label{fig:InO28}
\end{subfigure}
\caption{Binding energy discrepancies for (a) InH and (b) InO molecules using small-core [[Ar]3$d^{10}$]. Notation follows the same conventions as defined previously in Fig. \ref{fig:Rb_mols}.}
\end{figure*}

\subsubsection{Sb}
\begin{figure*}[!htbp]
\centering
\begin{subfigure}{0.5\textwidth}
\includegraphics[width=\textwidth]{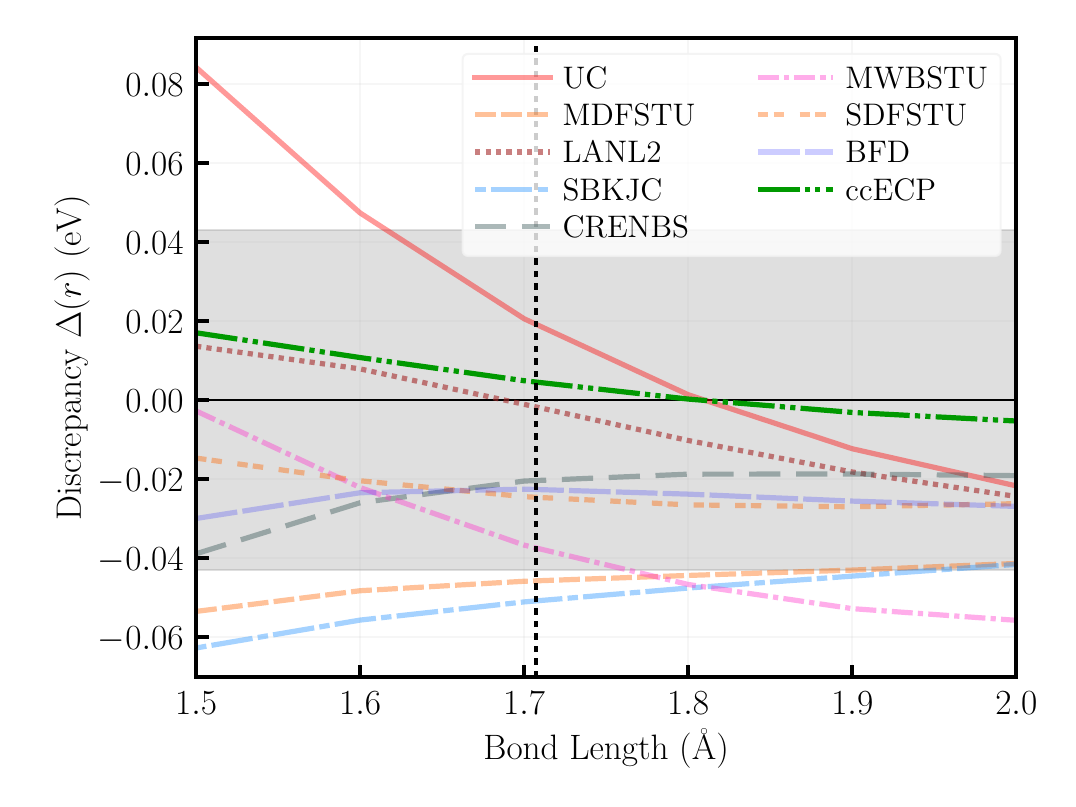}
\caption{SbH binding curve discrepancies}
\label{fig:SbH}
\end{subfigure}%
\begin{subfigure}{0.5\textwidth}
\includegraphics[width=\textwidth]{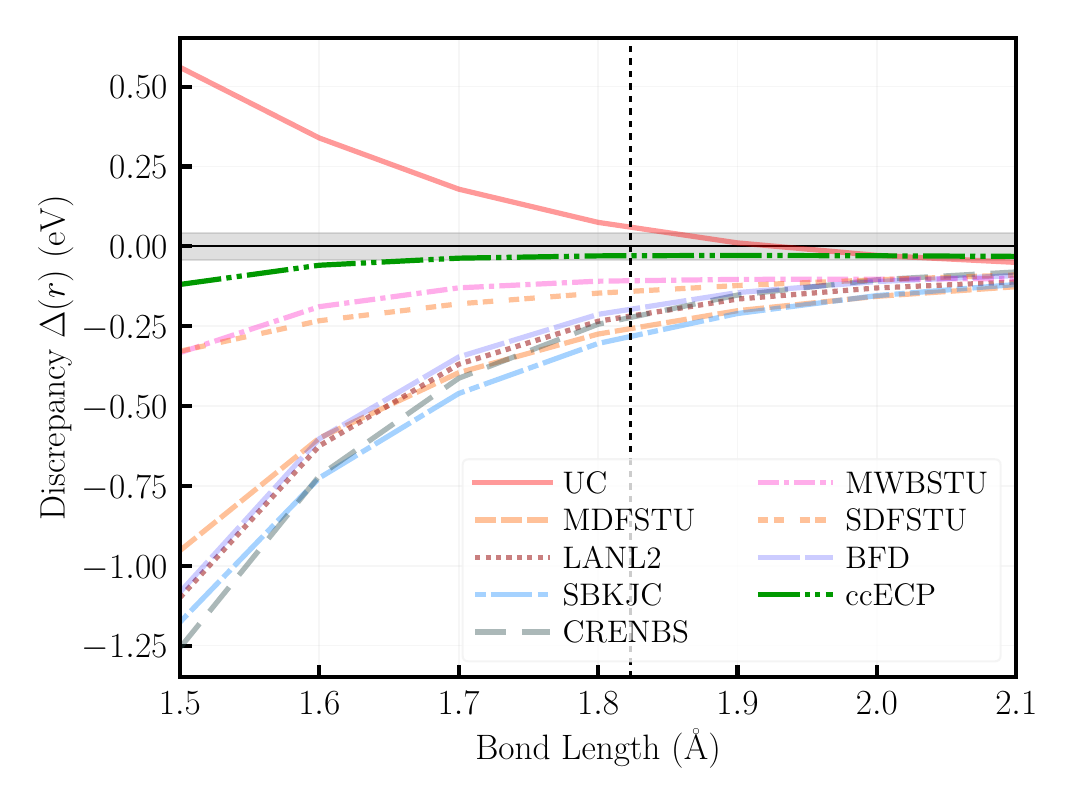}
\caption{SbO binding curve discrepancies}
\label{fig:SbO}
\end{subfigure}
\caption{Binding energy discrepancies for (a) SbH and (b) SbO molecules. Conventions are the same as defined previously in Fig. \ref{fig:Rb_mols}.} 
\label{fig:Sb_mols}
\end{figure*}

For antimony (Sb), there are nominally five valence electrons, making it rather difficult to capture accurate properties for all the tested possibilities. The results presented in Figs. \ref{fig:MAD_in_elements}-\ref{fig:WMAD_in_elements} indicate that our ccECP performs on par or better than most legacy ECPs and the UC method. 
While it is possible to further improve the MADs through optimization, it comes at the cost of negatively impacting the accuracy of the binding curve in SbO. The optimal compromise is encouraging, compared to the legacy constructions; see Fig. \ref{fig:Sb_mols}, while also providing lower MADs and LMADs. This required significant interventions in the optimization process of the ccECP parameters, 
in particular, special attention was given to the $s$ and local channels where we systematically refined the coefficients of these channels by explicitly varying the first and second digits after the decimal point. The optimizer was configured to constrain the specific exponents of the local channel, keeping them unchanged while allowing the remaining parameters to be optimized. These hand-tuned adjustments were carefully tracked to assess their impact on the SbO binding curve to determine which parameter eliminates overbinding as seen in the other comparing ECPs. This resulted in discrepancies that were well below the chemical accuracy near equilibrium bond lengths \ref{fig:Sb_mols}. 

\subsubsection{Pb}
\begin{figure*}[!htbp]
\centering
\begin{subfigure}{0.5\textwidth}
\includegraphics[width=\textwidth]{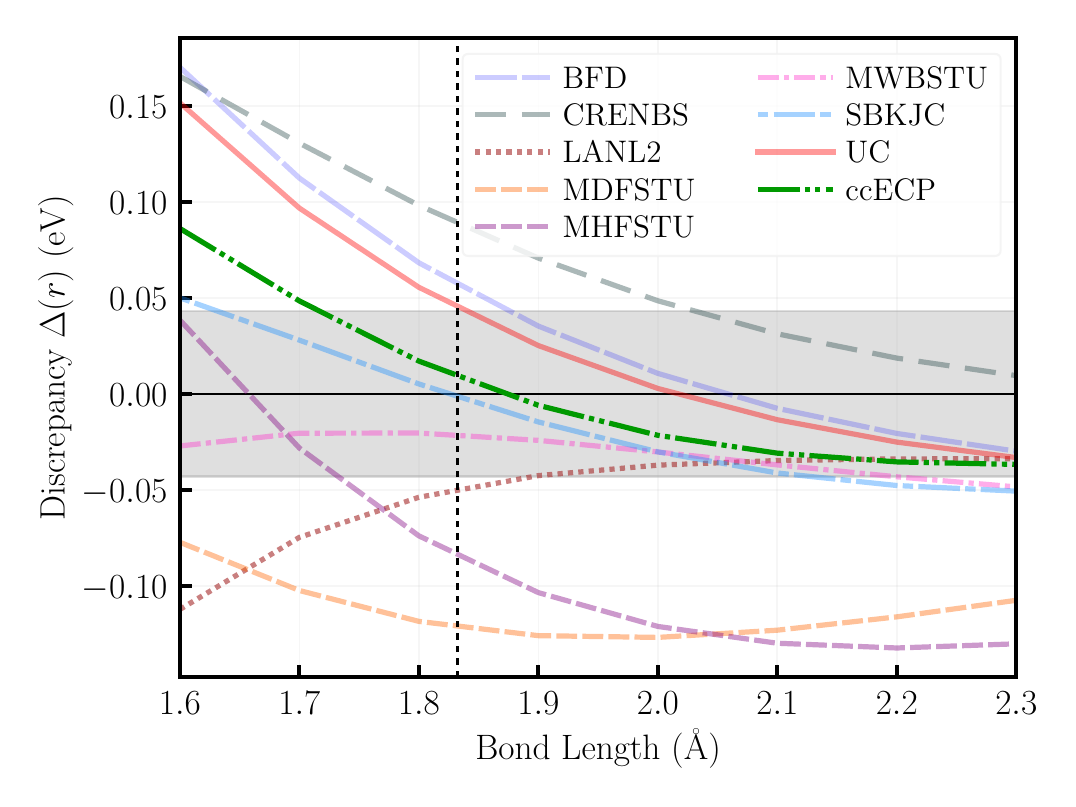}
\caption{PbH binding curve discrepancies}
\label{fig:PbH}
\end{subfigure}%
\begin{subfigure}{0.5\textwidth}
\includegraphics[width=\textwidth]{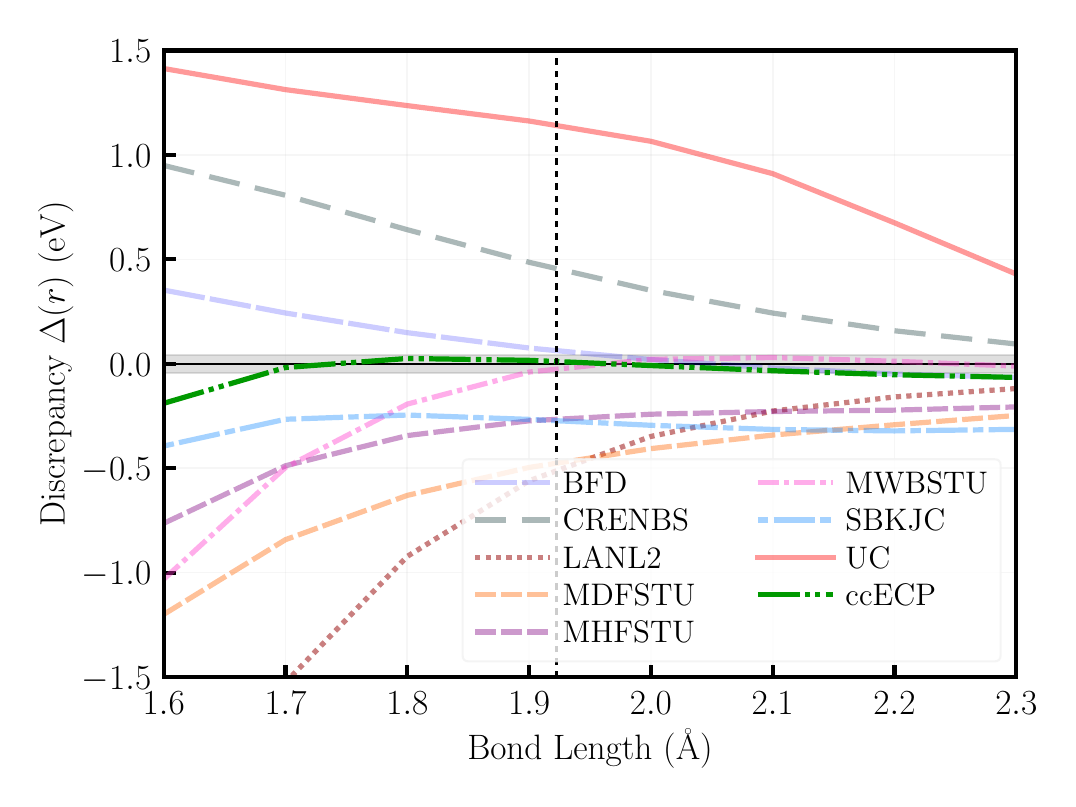}
\caption{PbO binding curve discrepancies}
\label{fig:PbO}
\end{subfigure}
\caption{Binding energy discrepancies for (a) PbH and (b) PbO molecules. Conventions are the same as defined previously in Fig. \ref{fig:Rb_mols}.}
\label{fig:Pb_mols}
\end{figure*}

For lead (Pb), converging calculations at the CCSD(T) level required the use of unrestricted CCSD(T) and adjustments of the energy denominators associated with single and double excitations and the application of the direct inversion of iterative subspace technique in order to aid the convergence.
In the atomic spectrum, the ccECP shows a significant improvement over legacy ECPs and the UC method, achieving the best overall metrics and approaching chemical accuracy.
Molecular calculations (Fig. \ref{fig:Pb_mols}) proved to be more challenging for ECPs, particularly for PbO, where legacy ECPs and the UC method exhibit discrepancies almost an order of magnitude or more beyond the chemical accuracy.
In PbH, our ccECP balances performance across bond lengths, addressing overbinding at the dissociation limit (seen in MDFSTU, MHFSTU, MWBSTU, and SBKJC) and underbinding at compressed bond lengths (observed with UC and CRENBL).
For PbO, the ccECP demonstrates consistent improvement, maintaining chemical accuracy across most geometries and at the equilibrium configuration, while significantly reducing errors at compressed bond lengths.

\subsection{Selected $4d$ elements}
For ruthenium (Ru) and cadmium (Cd) we use core [[Ar]3$d^{10}$], leaving 16 and 20 electrons in the valence space, respectively. 
\begin{table*}
\small
\centering
\caption{SOREP optimized parameters for the selected $4d$ elements ccECPs. The parameters follow the same definitions as in Table \ref{tab:selected_5s_params}.
}
\label{tab:selected_4d_params}
\begin{tabular}{cccccrrccccccrrr}
\hline\hline
\multicolumn{1}{c}{Atom} & \multicolumn{1}{c}{$Z_{\rm eff}$} & \multicolumn{1}{c}{Hamiltonian} & \multicolumn{1}{c}{$\ell$} & \multicolumn{1}{c}{$n_{\ell k}$} & \multicolumn{1}{c}{$\alpha_{\ell k}$} & \multicolumn{1}{c}{$\beta_{\ell k}$} & & \multicolumn{1}{c}{Atom} & \multicolumn{1}{c}{$Z_{\rm eff}$} & \multicolumn{1}{c}{Hamiltonian} & \multicolumn{1}{c}{$\ell$} & \multicolumn{1}{c}{$n_{\ell k}$} & \multicolumn{1}{c}{$\alpha_{\ell k}$} & \multicolumn{1}{c}{$\beta_{\ell k}$} \\
\hline
Ru & 16 & AREP &  0 & 2 &   11.528850 &  209.785831  && Cd & 20 & AREP &  0 & 2 &   13.621456 &  270.035778   \\
   &    &      &  0 & 2 &    5.091373 &   30.211243  &&    &    &      &  0 & 2 &    7.523456 &   38.856504   \\
   &    &      &  1 & 2 &   10.326472 &  146.247148  &&    &    &      &  1 & 2 &   12.674501 &  193.825168   \\
   &    &      &  1 & 2 &    4.575025 &   22.565456  &&    &    &      &  1 & 2 &    6.826303 &   31.857429   \\
   &    &      &  2 & 2 &    8.830547 &   66.402019  &&    &    &      &  2 & 2 &   10.715569 &   79.186521   \\
   &    &      &  2 & 2 &    3.215112 &    8.598081  &&    &    &      &  2 & 2 &    5.031907 &   12.823090   \\
   &    &      &  3 & 1 &   11.999981 &   16.000000  &&    &    &      &  3 & 1 &   10.991393 &   20.000000   \\
   &    &      &  3 & 3 &   12.003052 &  191.999691  &&    &    &      &  3 & 3 &   10.838913 &  219.827863   \\
   &    &      &  3 & 2 &   11.998013 &  -49.381269  &&    &    &      &  3 & 2 &   11.419750 & -124.388048   \\
   &    &      &  3 & 2 &    7.820339 &  -20.647860  &&    &    &      &  3 & 2 &   11.576062 &  -29.369023   \\
   &    &      &    &   &             &              &&    &    &      &    &   &             &               \\
   &    & SO   &  1 & 2 &   10.537904 &  -97.502222  &&    &    & SO   &  1 & 2 &   14.526780 & -129.083498   \\
   &    &      &  1 & 2 &   10.185064 &   97.496842  &&    &    &      &  1 & 2 &   10.434125 &  129.371140   \\
   &    &      &  1 & 2 &    4.743315 &  -15.718464  &&    &    &      &  1 & 2 &    5.094815 &  -21.247070   \\
   &    &      &  1 & 2 &    4.501084 &   15.331546  &&    &    &      &  1 & 2 &    8.573327 &   21.471063   \\
   &    &      &  2 & 2 &    8.847108 &  -26.972763  &&    &    &      &  2 & 2 &   11.255327 &  -31.612809   \\
   &    &      &  2 & 2 &    8.802027 &   26.950321  &&    &    &      &  2 & 2 &   11.039120 &   31.643819   \\
   &    &      &  2 & 2 &    3.207352 &   -3.352404  &&    &    &      &  2 & 2 &    4.827094 &   -4.984473   \\
   &    &      &  2 & 2 &    3.210690 &    3.488365  &&    &    &      &  2 & 2 &    4.686361 &    4.780917   \\
   &    &      &    &   &             &              &&    &    &      &    &   &             &              \\
\hline\hline
\end{tabular}
\end{table*}
\subsubsection{Ru}
\begin{figure*}[!htbp]
\centering
\begin{subfigure}{0.5\textwidth}
\includegraphics[width=\textwidth]{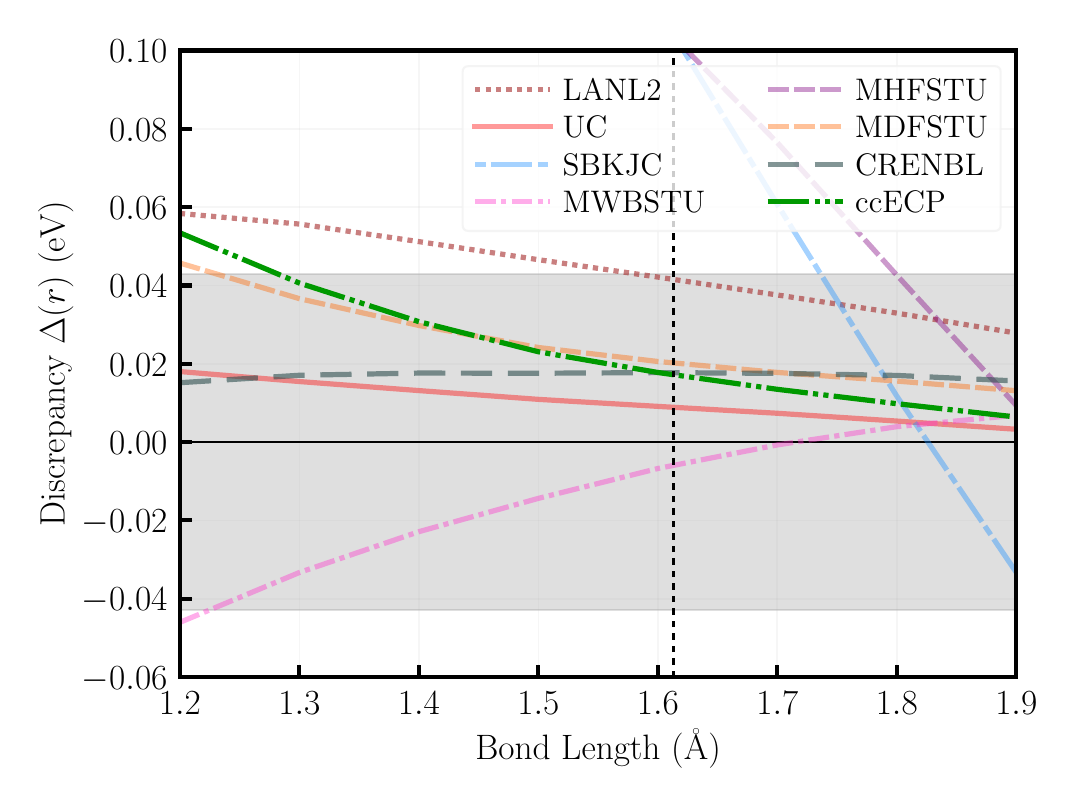}
\caption{RuH binding curve discrepancies}
\label{fig:RuH}
\end{subfigure}%
\begin{subfigure}{0.5\textwidth}
\includegraphics[width=\textwidth]{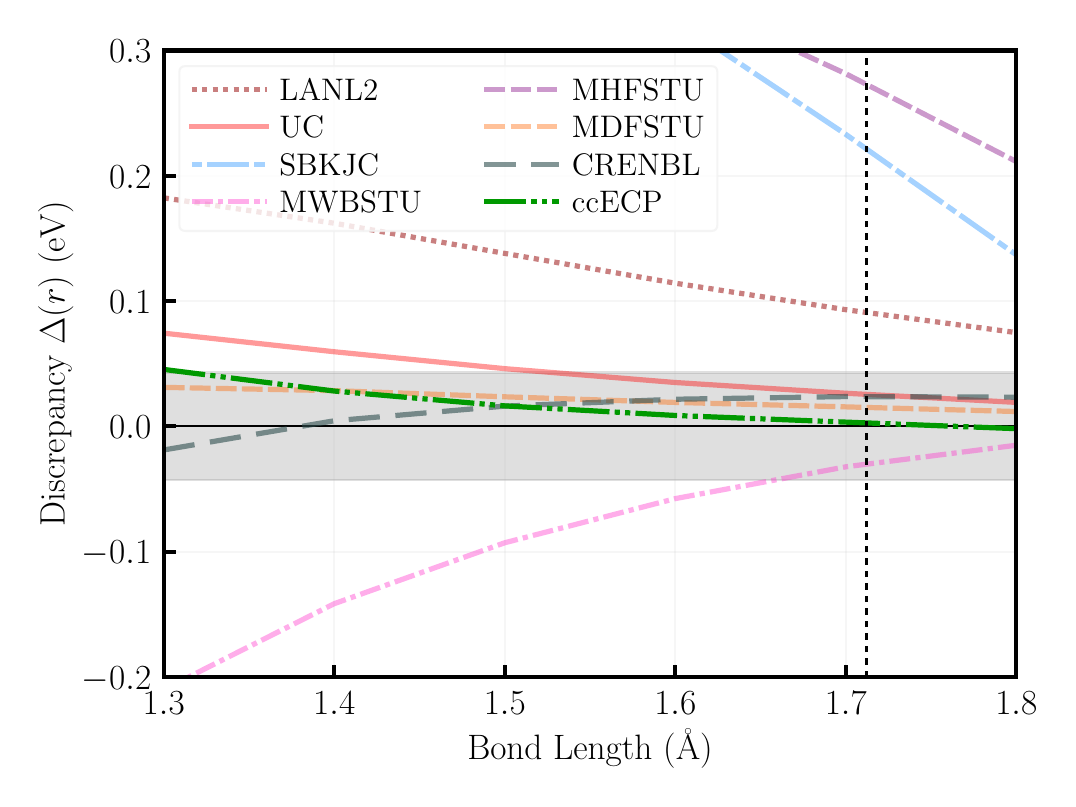}
\caption{RuO binding curve discrepancies}
\label{fig:RuO}
\end{subfigure}
\caption{Binding energy discrepancies for (a) RuH and (b) RuO molecules. Conventions are the same as defined previously in Fig. \ref{fig:Rb_mols}.}
\label{fig:Ru_mols}
\end{figure*}

Ru is an element that we have used in calculations before\cite{Gani-Ru2022}; however, in the past we simply smoothed out the Coulomb cusp present in the MDFSTU ECP.
We extended our efforts to optimize the full ccECP that performs comparably with that of CRENBL, which is the best performing legacy ECP in the atomic spectrum, with CRENBL slightly better in LMAD and ccECP slightly better in overall MAD.
The accuracy of the UC atomic spectrum is notably high, making it a challenging benchmark to match. In contrast, its performance in RuO (Fig. \ref{fig:RuO}) shows underbinding at compressed bond lengths, whereas the ccECP performs similarly to CRENBL and MDFSTU.
In RuH (Fig. \ref{fig:RuH}), ccECP performs within the chemical accuracy for most bond lengths and matches CRENBL and MDFSTU at the equilibrium bond length.
\subsubsection{Cd}
\begin{figure*}[!htbp]
\centering
\begin{subfigure}{0.5\textwidth}
\includegraphics[width=\textwidth]{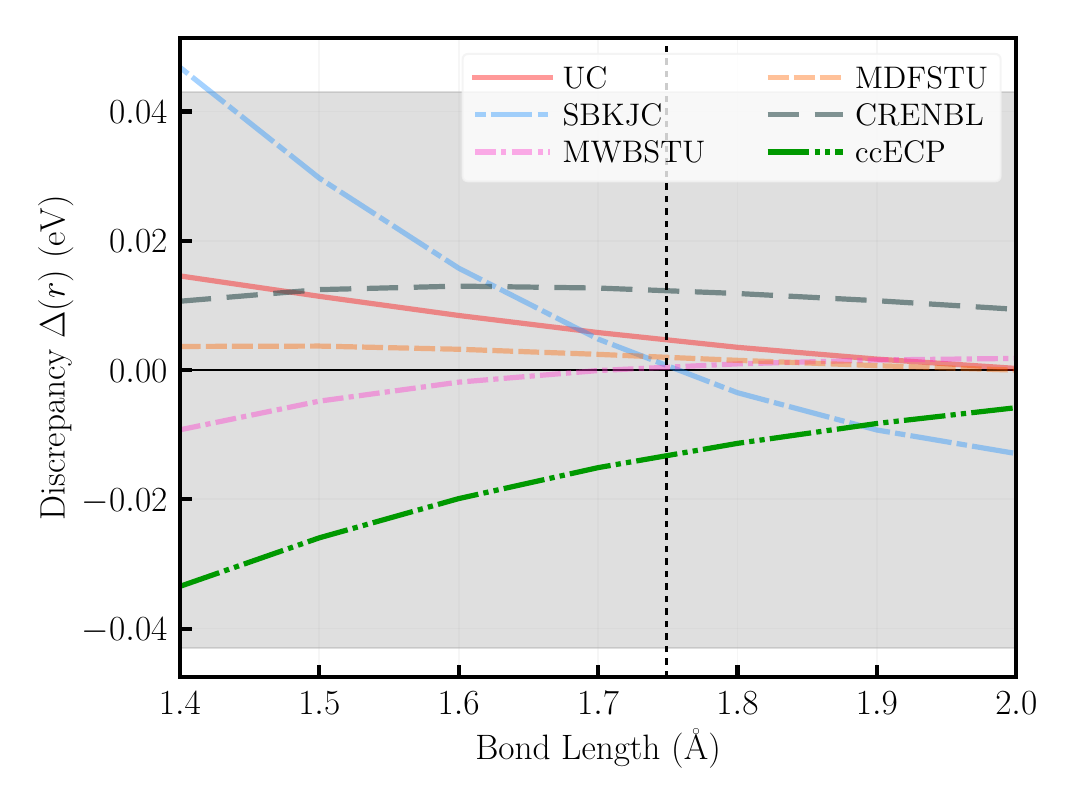}
\caption{CdH binding curve discrepancies}
\label{fig:CdH}
\end{subfigure}%
\begin{subfigure}{0.5\textwidth}
\includegraphics[width=\textwidth]{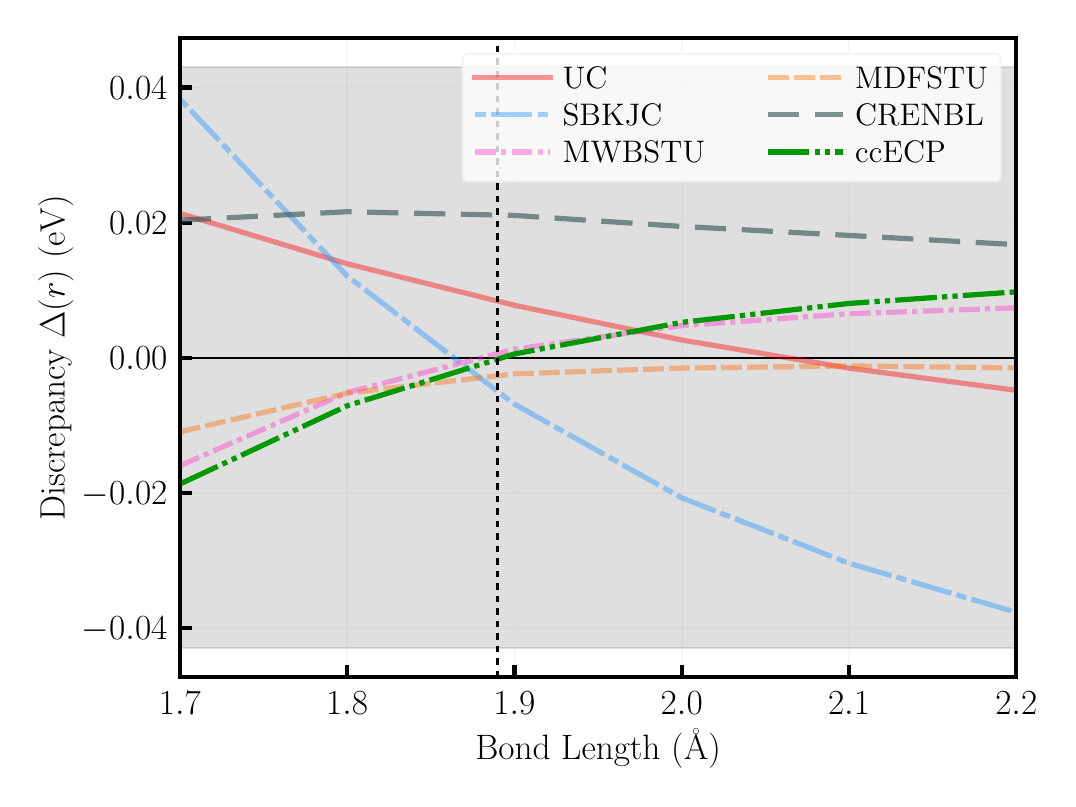}
\caption{CdO binding curve discrepancies}
\label{fig:CdO}
\end{subfigure}
\caption{Binding energy discrepancies for (a) CdH and (b) CdO molecules. Conventions are the same as defined previously in Fig. \ref{fig:Rb_mols}.}
\label{fig:Cd_mols}
\end{figure*}

As mentioned, the Cd core is [[Ar]3$d^{10}$] and that leaves 20 electrons in the valence space. This makes accurate construction relatively straightforward. The results of the first optimizations outperformed all legacy ECPs with the same cores for the atomic spectrum, and upon further investigation they also performed very well in the molecular binding tests (Fig. \ref{fig:Cd_mols}). The UC setting slightly outperforms our construction in terms of raw MAD, but the LMAD of ccECP moderately improves upon the UC. 
\subsection{Selected $4f$ elements}
\label{Selected 4f elements}
For the $4f$ ccECPs we use the same core choice as in our previous work that included selected lanthanide elements \cite{Haihan2024}, [[Kr]4$d^{10}$].
This core choice minimizes the number of electrons present in the valence space while at the same time accounting for the effects of $4f$ tails. This comes to the forefront in hybridizations of molecular or solid-state orbitals as well as in formation of large magnetic moments due to the partially filled $f$-shell.
For these ccECPs, we chose to reduce the number of highly ionized states (in the range +3 to +5) in optimization, as we noticed that their inclusion tended to unacceptably bias the overall quality of the ccECP in bonded environments.
We note that CRENBL and LANL2 had different cores for Ce and Eu, so they were not included in the tests.
\begin{table*}
\small
\centering
\caption{SOREP optimized parameters for the selected $4f$ elements ccECPs. The parameters follow the same definitions as in Table \ref{tab:selected_5s_params}.
}
\label{tab:selected_4f_params}
\begin{tabular}{cccccrrccccccrrr}
\hline\hline
\multicolumn{1}{c}{Atom} & \multicolumn{1}{c}{$Z_{\rm eff}$} & \multicolumn{1}{c}{Hamiltonian} & \multicolumn{1}{c}{$\ell$} & \multicolumn{1}{c}{$n_{\ell k}$} & \multicolumn{1}{c}{$\alpha_{\ell k}$} & \multicolumn{1}{c}{$\beta_{\ell k}$} & & \multicolumn{1}{c}{Atom} & \multicolumn{1}{c}{$Z_{\rm eff}$} & \multicolumn{1}{c}{Hamiltonian} & \multicolumn{1}{c}{$\ell$} & \multicolumn{1}{c}{$n_{\ell k}$} & \multicolumn{1}{c}{$\alpha_{\ell k}$} & \multicolumn{1}{c}{$\beta_{\ell k}$} \\
\hline
La & 11 & AREP &  0 & 2 &    3.484509 &   92.871354  && Ce & 12 & AREP &  0 & 2 &    4.002550 &  119.422791  \\
   &    &      &  0 & 2 &    4.565291 &   -2.537248  &&    &    &      &  0 & 2 &    2.091227 &   -2.011480  \\
   &    &      &  1 & 2 &    2.889889 &   64.778647  &&    &    &      &  1 & 2 &    3.164558 &   76.606896  \\
   &    &      &  1 & 2 &    2.506829 &    1.893740  &&    &    &      &  1 & 2 &    1.664888 &   -0.431335  \\
   &    &      &  2 & 2 &    2.018727 &   37.129096  &&    &    &      &  2 & 2 &    2.473670 &   68.864684  \\
   &    &      &  2 & 2 &    2.002775 &    1.372449  &&    &    &      &  2 & 2 &    2.275651 &   -3.747019  \\
   &    &      &  3 & 2 &    5.826438 &  -34.104154  &&    &    &      &  3 & 2 &    5.524740 &  -23.335057  \\
   &    &      &  3 & 2 &    4.749713 &    1.281856  &&    &    &      &  3 & 2 &    3.716386 &  -17.340566  \\
   &    &      &  4 & 1 &    8.000000 &   11.000000  &&    &    &      &  4 & 1 &    4.599500 &   12.000000  \\
   &    &      &  4 & 3 &    8.000000 &   88.000000  &&    &    &      &  4 & 3 &    4.599500 &   55.194000  \\
   &    &      &  4 & 2 &    6.000000 &  -70.000000  &&    &    &      &  4 & 2 &    4.600500 &  -54.149500  \\
   &    &      &  4 & 2 &    6.000000 &   -2.000000  &&    &    &      &  4 & 2 &    1.050500 &   -0.099500  \\
   &    &      &    &   &             &              &&    &    &      &    &   &             &              \\
   &    & SO   &  1 & 4 &    1.760495 &    2.848076  &&    &    & SO   &  1 & 4 &    2.174662 &    1.096988  \\
   &    &      &  2 & 4 &    4.112520 &   -0.382450  &&    &    &      &  1 & 4 &    2.101215 &    1.874880  \\
   &    &      &  3 & 4 &    2.966557 &    0.184712  &&    &    &      &  2 & 4 &    4.455950 &    0.390146  \\
   &    &      &    &   &             &              &&    &    &      &  2 & 4 &    2.528869 &    1.315403  \\
   &    &      &    &   &             &              &&    &    &      &  3 & 4 &    4.945406 &    0.112303  \\
   &    &      &    &   &             &              &&    &    &      &  3 & 4 &    3.247849 &   -0.012233  \\
Eu & 17 & AREP &  0 & 2 &    5.876250 &  250.008840  &&    &    &      &    &   &             &              \\
   &    &      &  0 & 2 &    3.840729 &    9.138778  &&    &    &      &    &   &             &              \\
   &    &      &  1 & 2 &   10.107840 &   13.515653  &&    &    &      &    &   &             &              \\
   &    &      &  1 & 2 &    2.849340 &   49.494929  &&    &    &      &    &   &             &              \\
   &    &      &  2 & 2 &    7.465373 &  249.551278  &&    &    &      &    &   &             &              \\
   &    &      &  2 & 2 &    1.323958 &    8.542164  &&    &    &      &    &   &             &              \\
   &    &      &  3 & 2 &    3.604610 &   -8.757847  &&    &    &      &    &   &             &              \\
   &    &      &  3 & 2 &   11.104512 &   11.191962  &&    &    &      &    &   &             &              \\
   &    &      &  4 & 1 &    4.815119 &   17.000000  &&    &    &      &    &   &             &              \\
   &    &      &  4 & 3 &    3.257988 &   81.857021  &&    &    &      &    &   &             &              \\
   &    &      &  4 & 2 &    3.506979 &  -87.860926  &&    &    &      &    &   &             &              \\
   &    &      &  4 & 2 &    0.810346 &   -0.758172  &&    &    &      &    &   &             &              \\
   &    &      &    &   &             &              &&    &    &      &    &   &             &              \\
   &    & SO   &  1 & 4 &    2.438722 &    4.446436  &&    &    &      &    &   &             &              \\
   &    &      &  1 & 4 &    2.134082 &   -2.047281  &&    &    &      &    &   &             &              \\
   &    &      &  2 & 4 &    2.604875 &    2.145353  &&    &    &      &    &   &             &              \\
   &    &      &  2 & 4 &    2.827415 &   -1.889994  &&    &    &      &    &   &             &              \\
   &    &      &  3 & 4 &    5.371059 &    2.201327  &&    &    &      &    &   &             &              \\
   &    &      &  3 & 4 &    5.168385 &   -1.839893  &&    &    &      &    &   &             &              \\
   &    &      &    &   &             &              &&    &    &      &    &   &             &              \\
   &    &      &    &   &             &              &&    &    &      &    &   &             &              \\
\hline\hline
\end{tabular}
\end{table*}
\subsubsection{La}
\begin{figure*}[!htbp]
\centering
\begin{subfigure}{0.5\textwidth}
\includegraphics[width=\textwidth]{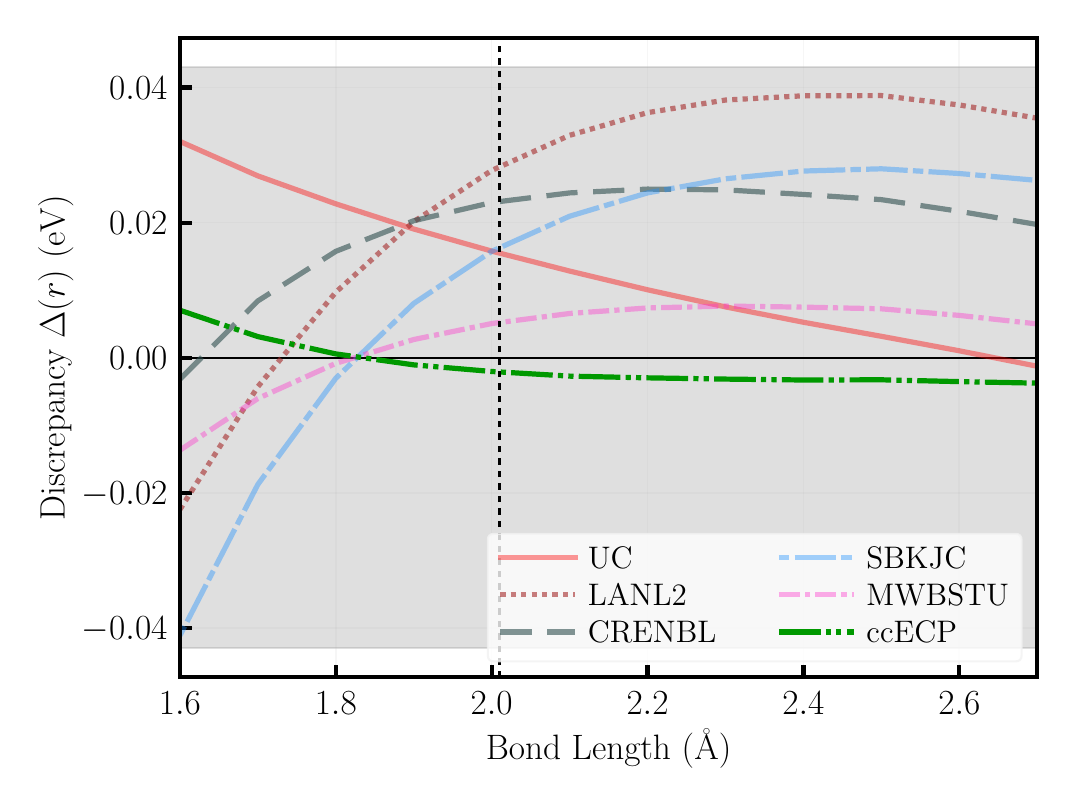 }
\caption{LaH binding curve discrepancies}
\label{fig:LaH}
\end{subfigure}%
\begin{subfigure}{0.5\textwidth}
\includegraphics[width=\textwidth]{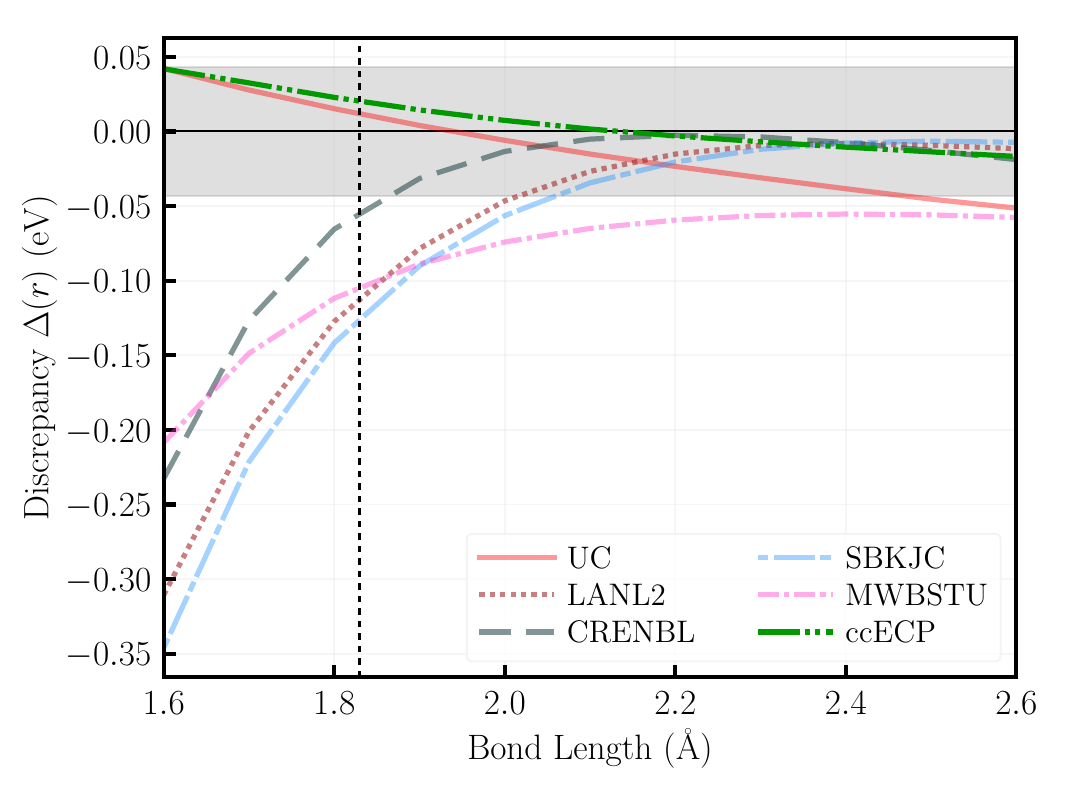}
\caption{LaO binding curve discrepancies}
\label{fig:LaO}
\end{subfigure}
\caption{Binding energy discrepancies for (a) LaH and (b) LaO molecules. Conventions are the same as defined previously in Fig. \ref{fig:Rb_mols}.}
\label{fig:La46_mols}
\end{figure*}

The ccECP for lanthanum (La) shows improved performance in atomic spectrum accuracy compared to the other legacy ECPs. We achieve the best values across all metrics, particularly in the LMAD. In LaH (Fig. \ref{fig:LaH}), where the electronic structure is simpler than in LaO (Fig. \ref{fig:LaO}), most of the ECPs perform within the chemical accuracy, of which the ccECP shows better consistency across the bond lengths tested, especially at equilibrium. 
For LaO, ccECP reduces the overbinding observed in legacy ECPs for compressed bond lengths, performing within chemical accuracy across all the geometries tested and showing better results than UC at the dissociation limit. 
During initial optimizations, we encountered challenges in converging states with partially filled $4f$ orbitals. We have addressed these issues by first optimizing the local channel and then the $f$ channel. Once satisfactory performance was achieved for these parameters, we followed the procedure outlined earlier in Sec. \ref{sec:optpro}.
%
\subsubsection{Ce}
\begin{figure*}[!htbp]
\centering
\begin{subfigure}{0.5\textwidth}
\includegraphics[width=\textwidth]{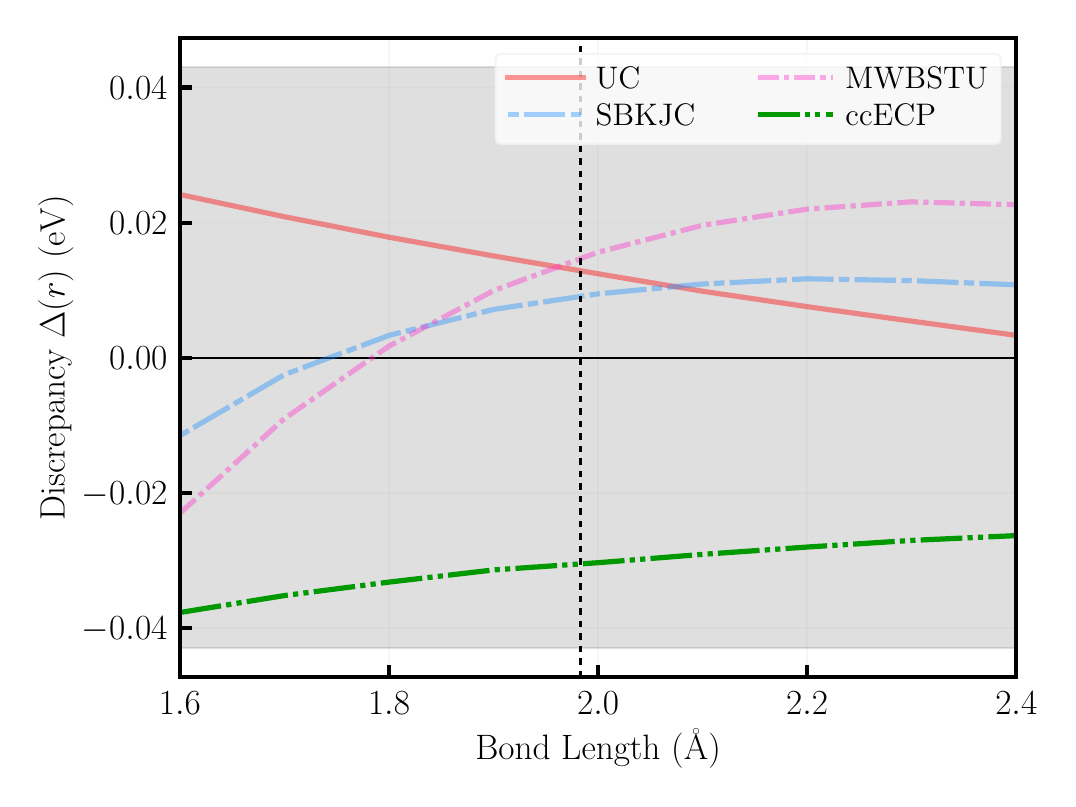}
\caption{CeH binding curve discrepancies}
\label{fig:CeH}
\end{subfigure}%
\begin{subfigure}{0.5\textwidth}
\includegraphics[width=\textwidth]{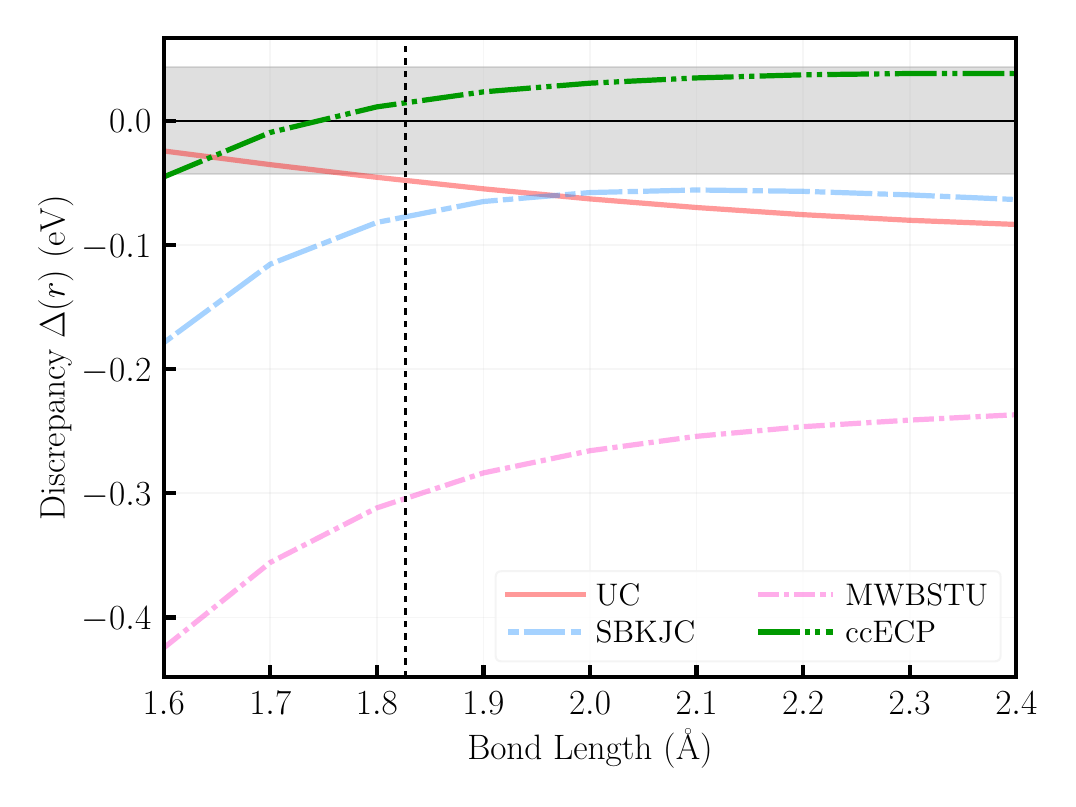}
\caption{CeO binding curve discrepancies}
\label{fig:CeO}
\end{subfigure}
\caption{Binding energy discrepancies for (a) CeH and (b) CeO molecules. Conventions are the same as defined previously in Fig. \ref{fig:Rb_mols}.}
\label{fig:Ce46_mols}
\end{figure*}

Cerium (Ce) lies second in the $4f$ series and has 2 partially filled orbitals, $4f^1$ and $5d^1$. 
This particular configuration posed several challenges in getting the ECP to perform within chemical accuracy in the molecular binding curves and the atomic spectrum simultaneously. During optimization, we observed that ECPs that performed extremely well for higher than 3+ ionized states were actually less optimal in molecules. Hence, for better transferability, we included ionized states up to 3+ 
as well as electron affinity and several excited states. With these considerations, we were able to construct a ccECP that outperforms the other legacy ECPs in the atomic spectrum (Figs. \ref{fig:MAD_in_elements}-\ref{fig:WMAD_in_elements}). 
In the case of molecular binding (Fig. \ref{fig:Ce46_mols}), particularly for CeO, the ccECP is the only one that lies within chemical accuracy at equilibrium and extended bond lengths, showing better agreement than the UC method and resides well within the required chemical accuracy for both molecules tested. 
This ensures that the ccECP has better transferability and applicability in bonding and solid-state environments without compromising on its performance for low-lying states in the atomic system.
%
\subsubsection{Eu}
\begin{figure*}[!htbp]
\centering
\begin{subfigure}{0.5\textwidth}
\includegraphics[width=\textwidth]{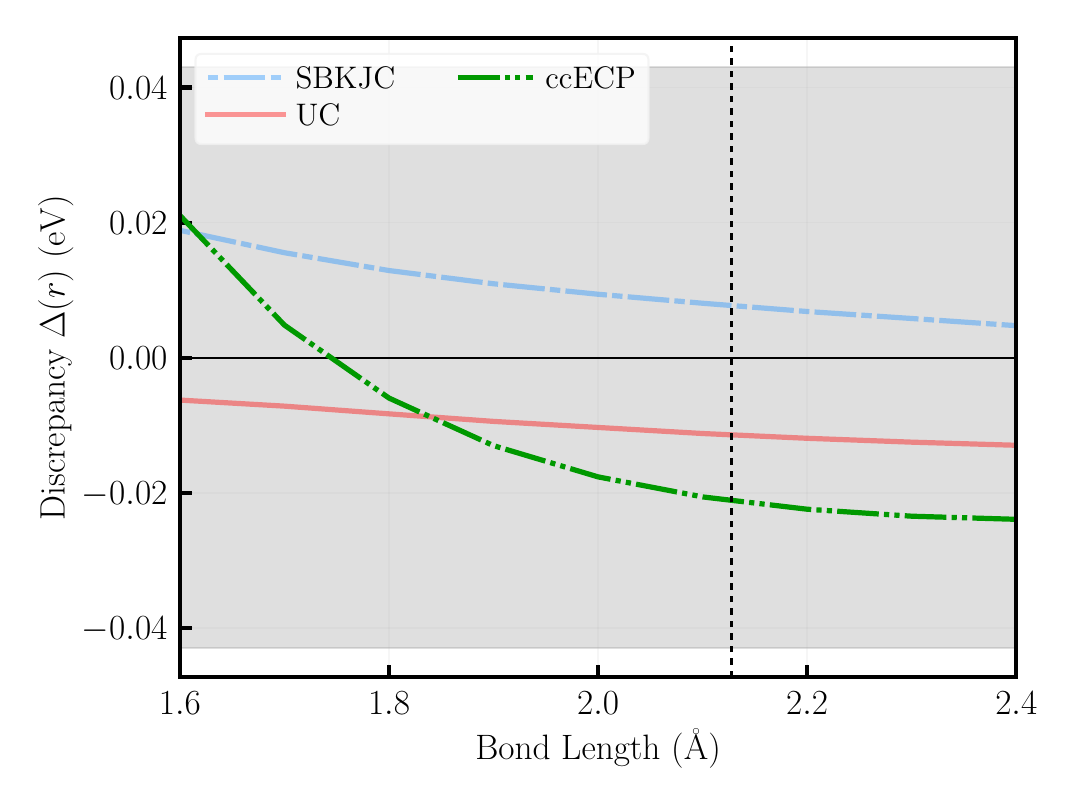 }
\caption{EuH binding curve discrepancies}
\label{fig:EuH}
\end{subfigure}%
\begin{subfigure}{0.5\textwidth}
\includegraphics[width=\textwidth]{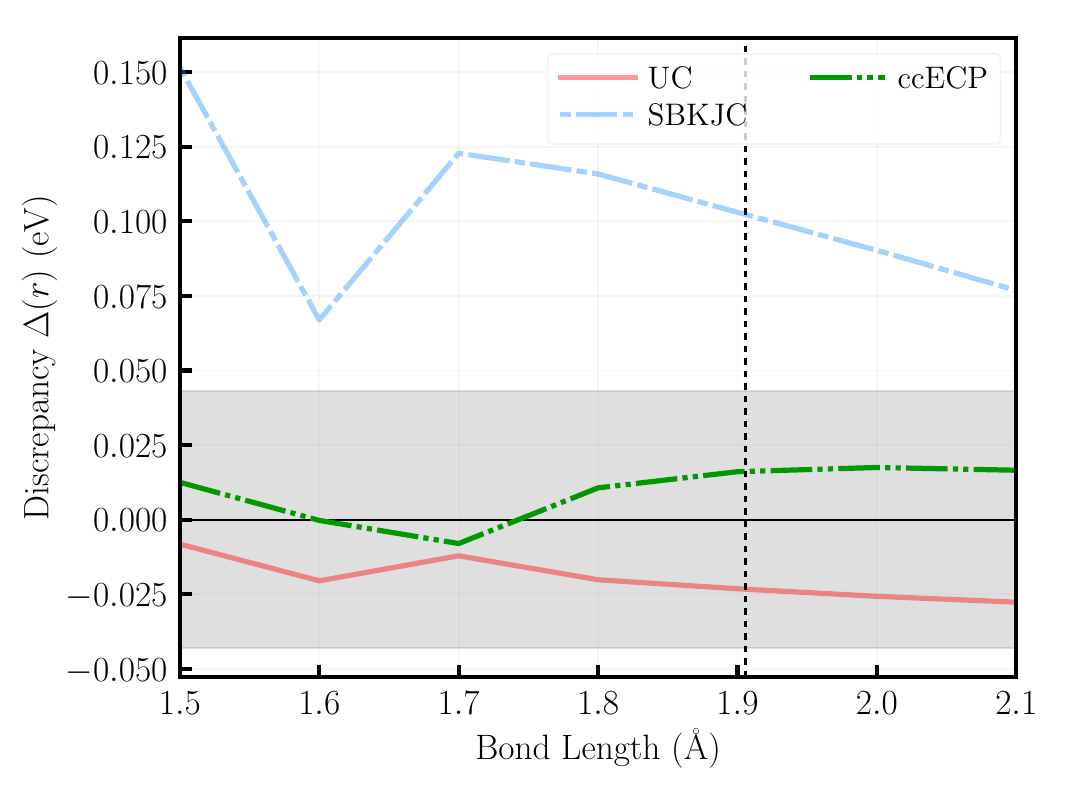}
\caption{EuO binding curve discrepancies}
\label{fig:EuO}
\end{subfigure}
\caption{Binding energy discrepancies for (a) EuH and (b) EuO molecules. EuO shows some signs of a change in state in the very compressed geometries, but the discrepancies do not look to substantially impact the trend of the rest of the data. Conventions are the same as defined previously in Fig. \ref{fig:Rb_mols}.}
\label{fig:Eu46_mols}
\end{figure*}

As with the previous $4f$ elements\cite{Haihan2024}, we crafted an initial guess by optimizing the non-local channels independently against $s$, $p$, $d$, and $f$ only states corresponding to which channel we were optimizing while a generalized local channel was left fixed. Later we optimized the local and non-locals both against the ground state and its deepest eigenvalues. When we preformed the final optimizations of the ccECP for europium (Eu), we reduced the number of highly ionized states used in the optimization, focusing on states with five or fewer ionizations and several excited states.
As seen with previous heavy elements, if the ECP is overfit to the atomic spectra, the performance in the molecules drops significantly.
We sought to strike a balance with performance in atomic and molecular contexts to maximize transferability.
As a result, the UC still outperforms the ccECP in the atomic spectrum on some fairly ionized states, but the ccECP is competitive on the low-lying states and in the molecules. 
SBKJC was the only other legacy ECP that shares the same core choice, and while it performs well in EuH, it struggles greatly with EuO, never entering the chemical accuracy regime.
EuO has some state crossings that occur in the more compressed region as seen in \ref{fig:EuO}. 
As a result, initially only geometries from $1.8$\AA\  to $2.1$\AA\  were converged, but with careful treatment, more compressed geometries were also resolved, although there is clearly a discrepancy in which the state was converged.
Despite this, judging by the trend in the figure, the discrepancies are small and show no indication that the performance of the ccECP in that region will change meaningfully.
%
%
%

\section{Conclusions}
\label{Conclusions}
We provide ccECPs for elements from the fifth and sixth rows including several from the $f$-block (Rb, Sr, Cs, Ba, In, Sb, Pb, Ru, Cd, La, Ce, and Eu). These elements were selected because of their importance for applications in catalysis, condensed matter physics and materials research.
For several elements (Sr and In), we have included small-core and large-core ccECP options in order to capture the effects of shallow cores states with large polarizabilities on electronic and chemical properties while still allowing large-scale computation with necessary accuracy.
As our constructions involve heavier elements, it becomes clear that although optimizing  against the AE atomic spectra is a good starting point for generating accurate ccECPs, its performance in chemical environments is dominated by the bonding and hybridization.
The behavior of an isolated atom and one in a molecule differs significantly in heavier elements due to the near-degeneracy of $4f$, $5d$, $6s$, and $6p$ levels, hence, the hybridization often involves the contribution of multiple one-particle atomic channels, which complicates the codependency of each channel.
In our constructions, careful attention was paid to achieving a balanced optimization of competing criteria, including transferability, fundamental accuracy, and the computational efficiency required for the construction to be worthwhile in a variety of high accuracy, correlated valence calculations.

\clearpage
\section*{Supplementary Material}

Additional information about ccECPs can be found in the Supplementary Material.
In particular, AREP results for calculated atomic spectra for each element and molecular binding curve fit parameters are provided.
The atomic spectra results include AE spectra for each element and the corresponding discrepancies for various core approximations.
Concerning the SO results, the detailed analysis of the J-splittings gaps at the COSCI level used for optimizing spin-orbit terms are listed as well.
The ccECPs in semi-local form, Kleinman-Bylander projected forms, as well as optimized Gaussian valence basis sets in various input formats (\textsc{Molpro}, \textsc{DIRAC}, \textsc{GAMESS}, \textsc{GAUSSIAN}, \textsc{NWChem}) can be found on the website \cite{pseudopotentiallibrary}.
Input and output data generated and related to this work will be published in Material Data Facility \cite{Blaiszik2016, Blaiszik2019} and can be found in Ref. \cite{mdf_data}.

\section*{acknowledgments}
The authors thank Paul R. C. Kent for reading the manuscript and providing helpful suggestions.
They are also grateful to Dr. Abdulgani Annaberdiyev for the fruitful discussions.
Axel Fraud helped with Cd AREP calculations. 

This work has been supported by the U.S. Department of Energy, Office of Science, Basic Energy Sciences, Materials Sciences and Engineering Division, as part of the Computational Materials Sciences Program and Center for Predictive Simulation of Functional Materials.

This research used resources of the National Energy Research Scientific Computing Center (NERSC), a U.S. Department of Energy Office of Science User Facility operated under Contract No. DE-AC02-05CH11231.

An award of computer time was provided by the Innovative and Novel Computational Impact on Theory and Experiment (INCITE) program.

This research used resources of the Oak Ridge Leadership Computing Facility, which is a DOE Office of Science User Facility supported under Contract No. DE-AC05-00OR22725.

This paper describes objective technical results and analysis. Any subjective views or opinions that might be expressed in the paper do not necessarily represent the views of the U.S. Department of Energy or the United States Government.

Note:  This manuscript has been authored by UT-Battelle, LLC, under contract DE-AC05-00OR22725 with the US Department of Energy (DOE). The US government retains and the publisher, by accepting the article for publication, acknowledges that the US government retains a nonexclusive, paid-up, irrevocable, worldwide license to publish or reproduce the published form of this manuscript, or allow others to do so, for US government purposes. DOE will provide public access to these results of federally sponsored research in accordance with the DOE Public Access Plan (http://energy.gov/downloads/doe-public-access-plan).

\section*{Conflict of Interest}
The authors have no conflicts to disclose.

\section*{Author Contributions}
\textbf{Omar Madany}: Conceptualization (lead); Investigation (lead); Methodology (equal); Visualization (equal); Writing – original draft (lead); Writing – review \& editing (lead).
\textbf{Benjamin Kincaid}: Conceptualization (equal); Investigation (equal); Methodology (lead); Visualization (equal); Writing – original draft (supporting); Writing – review \& editing (equal); Supervision (lead).
\textbf{Aqsa Shaikh}: Conceptualization (supporting); Investigation (equal); Methodology (supporting); Visualization ( supporting); Writing – original draft (supporting); Writing – review \& editing (equal).
\textbf{Elizabeth Morningstar}: Investigation (supporting).
\textbf{Lubos Mitas}: Conceptualization (equal); Investigation (supporting); Methodology (equal); Project administration (lead); Supervision (supporting); Writing – original draft (equal); Writing – review \& editing (equal).

\section*{DATA AVAILABILITY}
Data supporting the findings of this study are available in the article, its supplementary material, the pseudopotential library website\cite{pseudopotentiallibrary}, and the Material Data Facility\cite{mdf_data}.
.
\section*{REFERENCES}
\bibliographystyle{unsrt}
\bibliographystyle{apsrev4-2}
\bibliography{main.bib}
\newpage
\clearpage
\end{document}